\documentclass[pra,twocolumn]{revtex4-2}

\usepackage{amssymb}
\usepackage{graphicx}\usepackage{dcolumn}\usepackage{bm}\usepackage{physics}
\graphicspath{{result/}{./}}
\usepackage{float}
\usepackage{subfigure}
\usepackage{qcircuit}
\usepackage{hyperref}
\usepackage[usenames,dvipsnames]{xcolor}

\hypersetup{
    colorlinks,
    citecolor=blue,
    linkcolor=blue,
    urlcolor=blue
}

\usepackage[whole]{bxcjkjatype}
\usepackage[normalem]{ulem}
\usepackage{color}
\usepackage{soul}
\definecolor{americanrose}{rgb}{1.0, 0.01, 0.24}
\definecolor{electricpurple}{rgb}{0.75, 0.0, 1.0}
\definecolor{vividgreen}{rgb}{0.0, 0.6, 0.0}

\begin{document}

\preprint{APS/123-QED}

\title{Lie-Algebraic Subspace Quantization for Zero-Shot Quantum Learning and Barren-Plateau Mitigation}

\author{Yuhan Yao}
 \email{yao@biom.t.u-tokyo.ac.jp}
\author{Yoshihiko Hasegawa}\email{hasegawa@biom.t.u-tokyo.ac.jp}
\affiliation{Department of Information and Communication Engineering,
Graduate School of Information Science and Technology,
The University of Tokyo, Tokyo 113-8656, Japan
}\date{\today}

\begin{abstract}
The severe trainability bottlenecks inherent to variational optimization landscapes, particularly the barren plateau phenomenon, present a critical barrier to the practical scaling of parameterized quantum circuits (PQCs).
To circumvent the high cost of gradient-descent micro-tuning on quantum hardware, we introduce a rigorous, entirely analytical framework for zero-shot classical-to-quantum parameter mapping and manifold-based model merging.
We construct an analytical quantum compiler, the parameter transfer map, that converts a square classical neural-network weight into a low-dimensional subspace quantum evolution by combining Stiefel subspace selection, nearest-unitary polar projection, and logarithmic generator extraction.
The associated Subspace Quantization Theorem bounds the reconstruction error by two explicit contributions: geometric truncation and exact non-unitarity, with the latter equal to the retained block's singular-value deviation from the unitary manifold and becoming second order in the near-unitary pure-rotation regime.
Crucially, we show that when classical weights originate from identity-centered architectures such as residual networks, the local spectrum naturally breaks the uniform Haar measure distribution, maintaining a maximal topological distance from the matrix logarithm branch cut and providing initialization-time mitigation to the standard exponential gradient concentration.
For model fusion, transported generators are averaged on a covering frame, yielding an explicit error budget consisting of truncation loss, individual non-unitarity, and a second-order generator-separation term.
Separately, when the same construction is used as a warm-start initialization, the active dynamics are confined to a $k$-dimensional subspace, mitigating ambient barren-plateau scaling at initialization.
On IBM \texttt{ibm\_kobe}, the compiled circuits reach $0.987$ Hellinger fidelity at $k=8$, and subspace gradients remain resolvable up to $128$ physical qubits.
\end{abstract}

\maketitle

\section{Introduction\label{sec:introduction}}

Quantum information science and quantum computing have witnessed profound theoretical and experimental advancements, transitioning from foundational proofs-of-principle toward the era of noisy intermediate-scale quantum (NISQ) and early fault-tolerant devices \cite{endo2021nisq, bharti2022nisq, kjaergaard2020nisq}.
Despite the accelerating deployment of these processing platforms, their practical utility for executing complex, large-scale computational tasks remains severely constrained by hardware-level gate error rates \cite{cai2021quantum_error, cai2023quantum_error}, strict limitations on coherence qubits \cite{bergli2009limit_qubit, devoret2004limit_qubit}, and severe trainability bottlenecks inherent to variational optimization landscapes.
Chief among these optimization impediments is the barren plateaus phenomenon \cite{McClean2018, du2022bpvqa, cerezo2022bp, patti2021bp, shen2020bpqnn, wiersema2020bp}, where the gradients of typical cost functions concentrate exponentially around zero with the system size.
This gradient concentration renders standard gradient-based optimization protocols completely intractable, posing a fundamental barrier to the empirical viability of conventional parameterized quantum circuits (PQCs) \cite{tilly2022vqa, cerezo2021vqa} and quantum neural networks (QNNs) \cite{shen2020bpqnn}.

To circumvent the exponential trainability limitations of initializing PQCs from scratch, substantial collective effort has focused on establishing effective "warm-start" initialization protocols and quantum knowledge distillation strategies.
Recent paradigms leverage generative architectures or heuristic complex amplitude encoding schemes to seed variational quantum eigensolvers near optimal parameters \cite{Marrero2021, Grant2019, Cerezo2021, yao2025, yao2025directgradientcomputationbarren, yao2026gradientanalysisbarrenplateau}. Concurrently, advanced formulations analyze the dynamical Lie algebra (DLA) \cite{Goh2025DLA, Qva2025DLA, liang2026DLA, Ragone2024DLA, Wiersema2024DLA} underlying quantum circuits to elucidate the geometric properties of optimization landscapes or to compile compressed classical simulators for specific restricted multi-qubit systems. However, these established methodologies remain fundamentally limited; they either rely on empirical, heuristic fitting routines that offer no analytical guarantees, or they require costly gradient-descent optimization loops directly on the quantum hardware, which repeatedly exposes the system to gradient stagnation risks within deep landscape profiles.

In this paper, we address these challenges by introducing an entirely analytical framework for zero-shot classical-to-quantum parameter mapping and manifold-based model merging. Rather than treating the PQC as an unconstrained black box, we restrict the continuous quantum dynamics to a low-dimensional subspace geometry built from Stiefel frames and local unitary generators.
The resulting parameter transfer map takes a square neural-network weight, selects its dominant subspace by singular-value decomposition, projects the retained block to its nearest unitary by polar decomposition, and extracts a Hermitian generator by the principal matrix logarithm. This pipeline converts the executable, norm-preserving component of a classical layer into a compact subspace quantum evolution without fitting quantum parameters.
We prove a Subspace Quantization Theorem establishing that the reconstruction error is bounded by geometric truncation and exact non-unitarity, the latter being the retained block's singular-value deviation from the unitary manifold and becoming second order in the near-unitary regime.
The same construction provides an analytical warm-start for trainable quantum layers. Identity-centered residual architectures initialize the extracted generators near the identity of the local unitary group, avoiding the Haar-random regime underlying barren plateaus. A Weingarten-calculus analysis shows that the gradient variance depends on the chosen subspace dimension rather than the ambient qubit number.

The significance of this study lies in its potential to establish a training-free paradigm for hybrid quantum-classical computing.
By treating the Lie-algebraic representation as a structural communication protocol across the architectural boundary, our framework enables the immediate assembly of highly generalizable PQCs through purely classical, algebraic manipulation in the tangent space.
We further extend the construction to multi-model fusion by lifting each subspace generator to the ambient space, projecting it onto the covering subspace and restricting it back to a common Lie algebra.
In this shared frame, generator averaging yields a controlled merge whose additional nonlinearity is second order in the transported generator separation, while the remaining error is accounted for by coverage and non-unitarity terms.
This entirely eliminates the dependence on quantum-side gradient evaluation during initialization and model synthesis.
Beyond simulation, we validate the framework on the 156-qubit IBM Heron processor \texttt{ibm\_kobe}. The compiled subspace circuits retain high output fidelity for small active registers, achieving Hellinger fidelity $0.987$ at $k=8$ and $0.845$ at the four-qubit operating point $k=16$. A separate hardware trainability test further shows that the subspace-restricted gradient signal remains resolvable up to $128$ physical qubits, while a global hardware-efficient ansatz is pinned to the shot-noise floor.
Consequently, this work provides an analytical bridge between deep classical representations and fault-tolerant quantum embedding processors, offering an optimal, variational-free mechanism for scaling deep quantum machine learning architectures toward practical macroscopic applications.

\section{\label{sec:preliminaries}MATHEMATICAL AND GEOMETRIC PRELIMINARIES}

In this section, we establish the formal mathematical notation and the geometric framework that underpin our theoretical architecture.
We introduce the complex Stiefel manifold and its associated Lie algebra, define the hybrid operator ansatz, and rigorously delineate its subspace unitarity.
Furthermore, we characterize the intrinsic coordinate redundancy of this parameterization through the lens of gauge invariance and principal fiber bundles, providing a well-defined topological space for the parameter transfer maps developed in subsequent sections.

\subsection{\label{sec:sub_notation}Notation and Operator-Theoretic Setting}

Throughout this work, the ambient complex vector space representing the global quantum state space is denoted by $\mathbb{C}^N$, where $N$ represents the dimension of the Hilbert space.
For an $n$-qubit system, $N = 2^n$.
The standard Dirac inner product of two vectors $x, y \in \mathbb{C}^N$ is given by $\langle x, y \rangle = x^\dagger y$, where $\dagger$ denotes the conjugate transpose. 

The space of bounded linear operators on this Hilbert space is represented by the endomorphism algebra $\mathrm{End}(\mathbb{C}^N)$.
We equip $\mathrm{End}(\mathbb{C}^N)$ with the standard Hilbert-Schmidt (Frobenius) inner product, defined for any $A, B \in \mathrm{End}(\mathbb{C}^N)$ as $\langle A, B \rangle_F := \mathrm{Tr}(A^\dagger B)$.
This inner product naturally induces the Frobenius norm:
\begin{equation}
\|A\|_F = \sqrt{\mathrm{Tr}(A^\dagger A)}.
\end{equation}

\subsection{\label{sec:sub_manifolds}Complex Stiefel Manifolds and Subspace Projectors}

To implement a data-adaptive dimensionality reduction that preserves the kinematic properties of quantum states, we restrict our parameter space to a low-dimensional sub-manifold.
Let $k \ll N$ be the intrinsic dimension of the target subspace. 

The complex Stiefel manifold, denoted by $\mathrm{St}(k,N;\mathbb{C})$, is defined as the space of all orthonormal $k$-frames in $\mathbb{C}^N$:
\begin{equation}
\mathrm{St}(k,N;\mathbb{C}) := \left\{ Q \in \mathbb{C}^{N\times k} \ \middle|\ Q^\dagger Q = I_k \right\},
\end{equation}
where $I_k$ is the $k \times k$ identity matrix. 

Each point $Q \in \mathrm{St}(k,N;\mathbb{C})$ acts as a partial isometry embedding the local computational space $\mathbb{C}^k$ into the ambient space $\mathbb{C}^N$.
This embedding naturally induces a unique Hermitian orthogonal projection operator onto the subspace $\mathcal{H}_k = \mathrm{span}(Q)$:
\begin{equation}
P_Q := Q Q^\dagger,
\end{equation}
which satisfies the algebraic properties of idempotency ($P_Q^2 = P_Q$) and self-adjointness ($P_Q^\dagger = P_Q$).

% =====================================================================
\subsection{\label{sec:sub_lie_theory}Unitary Lie Groups and Lie Algebras}

To formalize the continuous rotations within the computational subspace, we briefly review the relevant elements of Lie theory. In quantum mechanics, the set of all norm-preserving linear transformations on a $k$-dimensional complex Hilbert space $\mathcal{H}_k \cong \mathbb{C}^k$ forms the unitary group $U(k)$, defined as:
\begin{equation}
    U(k) := \{ U \in \mathrm{End}(\mathbb{C}^k) \mid U^\dagger U = U U^\dagger = I_k \}.
\end{equation}
The unitary group $U(k)$ is a compact, connected real Lie manifold of dimension $k^2$. 

The identity component of this manifold is generated by its tangent space at the identity, known as the Lie algebra $\mathfrak{u}(k)$.
Structurally, $\mathfrak{u}(k)$ consists of all $k \times k$ skew-Hermitian matrices:
\begin{equation}
    \mathfrak{u}(k) := \{ X \in \mathrm{End}(\mathbb{C}^k) \mid X^\dagger = -X \}.
\end{equation}
In physics, it is conventional to express elements of the Lie algebra via self-adjoint (Hermitian) operators. Any skew-Hermitian matrix $X \in \mathfrak{u}(k)$ can be written as $X = -iH$, where $H \in \mathrm{End}(\mathbb{C}^k)$ is a strictly Hermitian physical Hamiltonian satisfying $H^\dagger = H$.

The topological bridge between the vector space of the Lie algebra $\mathfrak{u}(k)$ and the curved manifold of the Lie group $U(k)$ is established by the matrix exponential map, $\exp: \mathfrak{u}(k) \to U(k)$, defined via its universally convergent Taylor series:
\begin{equation}\label{eq:exp_map_def}
    \exp(X) = e^{X} = \sum_{m=0}^\infty \frac{X^m}{m!}.
\end{equation}
For the compact unitary group $U(k)$, every unitary matrix admits at least one skew-Hermitian logarithm, so the exponential map from $\mathfrak{u}(k)$ onto $U(k)$ is surjective. 

Conversely, within a localized topological neighborhood around the identity operator $I_k$, we can define a smooth inverse mapping via the principal matrix logarithm, $\log: U(k) \to \mathfrak{u}(k)$. For a unitary matrix $U \in U(k)$ whose spectrum $\sigma(U)$ does not intersect the negative real axis $(-\infty, 0]$, the principal logarithm provides a unique, differentiable vector field representation in the tangent space.
Within this branch-cut-free neighborhood, the subspace quantum evolution is uniquely parameterized by the principal Hermitian generator, providing the geometric foundation for the parameter transfer maps implemented in Sec.~\ref{sec:main_theory}.

\subsection{\label{sec:sub_hybrid_ansatz}The Hybrid Operator and Subspace Unitarity}

To formalize the localized quantum dynamics, we introduce a hybrid algebraic structure that couples embedding geometries with subspace evolutions. 
Given a frame-generator pair $(Q,H) \in \mathrm{St}(k,N;\mathbb{C}) \times i\,\mathfrak{u}(k)$,
where $H = H^\dagger$ is Hermitian so that $-iH \in \mathfrak{u}(k)$ and $e^{-iH}\in U(k)$,
we define the hybrid operator $\mathcal{O}(Q,H) \in \mathrm{End}(\mathbb{C}^N)$ as:
\begin{equation}\label{eq:hybrid_def}
\mathcal{O}(Q,H) := Q e^{-iH} Q^\dagger.
\end{equation}

This hybrid construction exhibits specific geometric constraints 
and admits an equivalent global representation, as stated below.

\textit{Lemma 1 (Subspace Unitarity and Global Projections).---}%
The hybrid operator $\mathcal{O}(Q,H)$ is a partial isometry that restricts to a strictly unitary evolution within $\mathrm{span}(Q)$, and admits an equivalent dual-projection representation:
\begin{equation}\label{eq:prop_unitarity}
\left.\mathcal{O}(Q,H)\right|_{\mathrm{span}(Q)} \in U(k),
\end{equation}
and
\begin{equation}\label{eq:prop_global_exp}
\mathcal{O}(Q,H) = P_Q e^{-i\tilde H} P_Q,
\end{equation}
where $\tilde H := Q H Q^\dagger \in \mathrm{End}(\mathbb{C}^N)$ represents the globally embedded Hamiltonian.

Geometrically, the forward map $\mathcal{O}(Q,H)$ constructs a three-step dynamic mechanism: $Q^\dagger$ projects and maps an extensive ambient state down to a $k$-dimensional computational core, $e^{-iH}$ implements a coherent quantum evolution within this subspace, and $Q$ lifts the state back into the global Hilbert space.
Lemma~1 guarantees the mathematical equivalence between this local flow and a globally projected exponential. 
Crucially, Eq.~\eqref{eq:prop_global_exp} reveals that executing this dynamic layer does not require the exponentiation of a full $N \times N$ matrix, providing a fundamental mechanism for the efficient classical simulation of localized quantum embeddings.
The detailed algebraic verification of these geometric properties is deferred to Appendix~\ref{app:proofs}.

\subsection{\label{sec:sub_bundle}Gauge Structure and Associated Bundles}

The parameterization of the hybrid operator via the pair $(Q,H)$ contains an implicit coordinate redundancy, which corresponds to the choice of an orthonormal basis within the subspace. This redundancy can be formalized as an internal structural symmetry under the action of the unitary Lie group $U(k)$.
Specifically, two parameter pairs $(Q,H)$ and $(Q',H')$ are physically equivalent, denoted by $(Q,H) \sim (Q',H')$, if they are connected by a gauge transformation matrix $R \in U(k)$ such that $Q' = QR$ and $H' = R^\dagger H R$. 

This coordinate artifice does not affect the physical transition amplitude, leading to the following core geometric property of our ansatz.

\textit{Theorem 1 (Gauge Invariance and Bundle Mapping).---}%
The hybrid evolution operator $\mathcal{O}(Q,H)$ is strictly invariant under the action of the local gauge group $U(k)$, satisfies
\begin{equation}\label{eq:gauge_inv}
    \mathcal{O}(QR, R^\dagger H R) = \mathcal{O}(Q,H), \quad \forall R \in U(k),
\end{equation}
and projects uniquely onto the quotient space under the equivalence relation $\sim$.
Consequently, the assignment defines a gauge-invariant map that is well defined on the associated bundle:
\begin{equation}\label{eq:bundle_map}
    \mathcal{O}: \mathrm{St}(k,N;\mathbb{C}) \times_{U(k)} i\,\mathfrak{u}(k) \longrightarrow \mathrm{End}(\mathbb{C}^N).
\end{equation}

Geometrically, a naive formulation would place the domain of $\mathcal{O}$ on the standard Cartesian product $\mathrm{St}(k,N;\mathbb{C}) \times i\,\mathfrak{u}(k)$. 
However, Eq.~\eqref{eq:gauge_inv} demonstrates that the directional fibers along the gauge orbits map onto identical physical operators.
By taking the quotient with respect to the group action, we construct the associated fiber bundle $\mathrm{St}(k,N;\mathbb{C}) \times_{U(k)} i\,\mathfrak{u}(k)$ over the base space defined by the complex Grassmannian manifold $\mathrm{Gr}(k,N;\mathbb{C}) \cong \mathrm{St}(k,N;\mathbb{C})/U(k)$. 
This quotient operation eliminates the internal basis degrees of freedom, structurally establishing the hybrid ansatz as a well-defined physical coordinate-free observable. 
The detailed algebraic verification of this gauge symmetry is presented in Appendix~\ref{app:gauge_proof}.

% =====================================================================
\subsection{\label{sec:sub_bp_immunity}Trainability: Polynomial Gradients via Subspace Warm-Start Initialization}

Having established the structural approximation boundaries of the hybrid quantum ansatz, we now address its dynamical trainability.
A notorious bottleneck in standard parameterized quantum circuits (PQCs) is the barren plateau phenomenon, wherein the gradients of typical cost functions concentrate exponentially around zero, rendering classical optimization concentration-limited.
We show that the transfer map of this work acts as an analytical warm-start initialization: it confines the trainable dynamics to a fixed low-dimensional Stiefel sub-manifold and seeds them in the identity neighborhood, so that gradient magnitudes scale polynomially in the subspace dimension $k$ rather than exponentially in the qubit number $n$.

Consider a global Hilbert space $\mathcal{H}_N \cong \mathbb{C}^N$ associated with an $n$-qubit system ($N = 2^n$). Let $Q \in \mathrm{St}(k, N; \mathbb{C})$ denote the isometric frame of truncation rank $k$.
We define the parameterized subspace-restricted quantum evolution operator as 
\begin{equation}
U(\boldsymbol{\theta})
:= Q e^{-i\left(H_0 + \sum_{j} \theta_j V_j\right)} Q^\dagger
+ \left(I_N - Q Q^\dagger\right),
\end{equation}
where $H_0$ is the analytical initial generator extracted by the transfer map and the $V_j=V_j^\dagger$ are localized Hermitian parametric generators.
The complementary identity term extends the subspace evolution to a unitary operator on the full Hilbert space: the first term acts as the trainable evolution on $\mathrm{span}(Q)$, while $I_N-QQ^\dagger$ acts trivially on $\mathrm{span}(Q)^\perp$.
For the subspace-supported initial states considered here, this complementary component drops out of the reduced dynamics but guarantees $U(\boldsymbol{\theta})^\dagger U(\boldsymbol{\theta})=I_N$.
Given an initial subspace-supported state $|\Psi_0\rangle = Q |x\rangle$ ($\|x\|_2 = 1$) and a global observable $O \in \mathrm{End}(\mathcal{H}_N)$, the cost function is $\mathcal{L}(\boldsymbol{\theta}) = \langle \Psi_0 | U^\dagger(\boldsymbol{\theta}) O U(\boldsymbol{\theta}) | \Psi_0 \rangle$.
The active dynamics live on a register of only $m = \log_2 k$ qubits, and the gradient variance is governed by $k$ alone, as stated below.

\textit{Theorem 2 (Polynomial Gradient Floor).---}%
Even in the worst case in which the local evolution $\tilde U(\boldsymbol\theta)=e^{-i(H_0+\sum_j\theta_jV_j)}$
forms a unitary $2$-design over the subspace group $U(k)$, the variance of the cost-function
gradient with respect to any parameter $\theta_j$ is
\begin{equation}\label{eq:bp_variance_bound}
    \mathrm{Var}\left[ \frac{\partial \mathcal{L}}{\partial \theta_j} \right]
    = \frac{\big\|[\tilde{O}, V_j]\big\|_F^2}{k+1} \sim \Omega\!\left(\frac{1}{k}\right),
\end{equation}
where $\tilde{O} = Q^\dagger O Q \in \mathrm{End}(\mathbb{C}^k)$ is the effective projection of the global observable onto the subspace.

Eq.~\eqref{eq:bp_variance_bound} is the crux of the trainability argument.
A generic PQC that explores the full $2^n$-dimensional Hilbert space has its integration measure tied to the ambient dimension, so its gradient variance decays exponentially with the ambient qubit number, as observed below by the $2^{-n}$ reference scaling.
Our construction instead reduces the relevant integral to the compact group $U(k)$ on a fixed $m=\log_2 k$-qubit register (Appendix~\ref{app:bp_proof}), so the variance floor is polynomial in the subspace dimension and independent of $n$.
Because $k$ is a classical SVD hyperparameter that does not grow with the nominal qubit count, the warm-start converts the gradient scaling from exponential to polynomial.

This bound is a worst-case floor; the analytical initialization is strictly better.
Because the hybrid operators are compiled from a pre-trained residual network initialized near the identity ($W \approx I_N$), the transfer map yields $\|H_0\|_F \ll 1$, so at $\boldsymbol\theta=0$ the evolution $\tilde U$ sits in the identity neighborhood of $U(k)$, which is the opposite of a $2$-design, where $\partial_j\mathcal{L}\big|_0 \approx i\langle x|[V_j,\tilde O]|x\rangle = \mathcal{O}(1)$.
We emphasize the scope of the claim: Eq.~\eqref{eq:bp_variance_bound} characterizes the initialization, guaranteeing a non-flat starting point rather than the absence of plateaus along the entire optimization trajectory.
In the zero-shot regime studied below the generators are frozen ($\boldsymbol\theta$ absent); opening the $V_j$ for few-shot fine-tuning is the trainable extension to which the bound applies.
The full Weingarten derivation is given in Appendix~\ref{app:bp_proof}.

\section{\label{sec:main_theory}QUANTUM PARAMETER ISOMORPHISM AND SUBSPACE QUANTIZATION}

In this section, we present the core methodological contribution of this work: an exact, analytical parameter transfer map that constructs an analytical subspace quantization map from near-unitary classical neural-network weights to subspace-restricted quantum generators.
We first define the rigorous operational pipeline of the transfer map via a joint singular value and polar decomposition.
We then introduce the spectral conditions required for topological stability.
Finally, we formulate and prove the Subspace Quantization Theorem, providing a strict micro-perturbative bound on the geometric approximation error.

% =====================================================================

\subsection{\label{sec:sub_transfer_map}The Parameter Transfer Map}

A foundational challenge in integrating pre-trained classical networks with quantum co-processors is the non-unitary nature of classical operator weights. 
General weight matrices $W \in \mathrm{End}(\mathbb{C}^N)$ typically exhibit full numerical rank, possess non-trivial singular-value spectra, and lack Hermiticity. 
To anchor these operators within a parameterized quantum circuit (PQC) without relying on variational micro-tuning, we introduce an analytical inverse mapping that projects classical information directly onto quantum Lie algebras.

\begin{figure}[h]
    \centering
    \includegraphics[width=\linewidth]{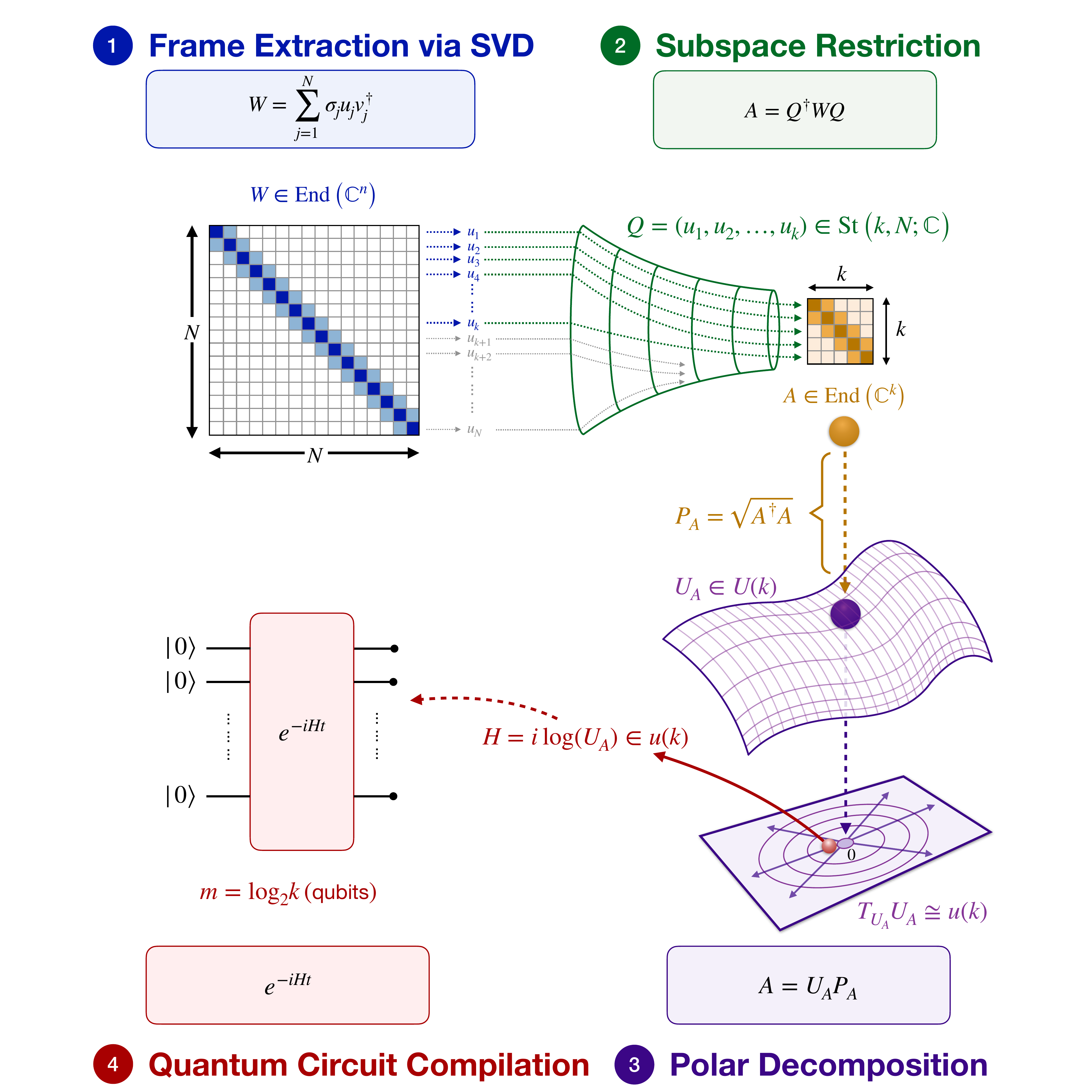}
    \caption{\textbf{Algebraic architecture and geometric projection mechanics of the quantum parameter transfer map $\mathcal{T}$ (Theorem~3).} 
    The conceptual workflow translates an arbitrary, non-unitary classical prior into a physically realizable subspace quantum evolution via four explicit algebraic operations: 
    \textbf{(1) Spectral Decoupling:} The ambient operator $W \in \mathrm{End}(\mathbb{C}^N)$ is transformed via singular value decomposition (SVD) to isolate the dominant dynamical sectors characterized by the leading $k$ left singular vectors. 
    \textbf{(2) Isometric Compression:} The isometric frame $Q \in \mathrm{St}(k, N; \mathbb{C})$ restricts the ambient interactions onto a $k$-dimensional computational core, yielding the localized effective operator $A$. 
    \textbf{(3) Manifold Quantization:} The non-unitary leakage $P_A = \sqrt{A^\dagger A}$ is geometrically isolated as the orthogonal distance (vertical projection error) between the core $A$ and the unitary manifold $U(k)$, distilling the optimal norm-preserving subspace rotation $U_A$. Concurrently, the tangent space $T_{U_A}U(k) \cong \mathfrak{u}(k)$ is invoked at the projection locus. 
    \textbf{(4) Logarithmic Lie Projection:} The principal matrix logarithm $\log(U_A)$ solves the inverse time-independent Schrödinger equation within the localized sector, analytically mapping the state from the group identity neighborhood (origin $0$) onto the exact physical Hermitian generator (Hamiltonian) $H$. The resulting observable specifies the quantum circuit layer $e^{-iHt}$ on an $m=\log_2 k$ active qubit register, mitigating barren-plateau scaling through explicit dimension reduction.}
    \label{fig:quantum_parameter_transfer}
\end{figure}

\textit{Theorem 3 (Quantum Parameter Transfer Pipeline).---}%
Given a classical weight matrix $W$ and a chosen top-$k$ SVD subspace whose retained
block $A$ is nonsingular and whose polar factor satisfies the branch-cut assumption~\eqref{eq:branch_cut}, the map
extracts a frame-generator pair $(Q,H)$, unique up to the $U(k)$ gauge freedom of
the retained subspace, as shown in Fig.~\ref{fig:quantum_parameter_transfer}.
\begin{enumerate}
    \item \textit{Frame Extraction via SVD:} Compute the singular value decomposition $W = \sum_{j=1}^N \sigma_j u_j v_j^\dagger$, and construct the isometric frame $Q \in \mathrm{St}(k,N;\mathbb{C})$ from the leading $k$ left singular vectors, namely $Q = (u_1, u_2, \dots, u_k)$.
    
    \item \textit{Subspace Restriction:} Compress the global ambient operator $W$ onto the $k$-dimensional computational core, yielding the localized matrix 
    \begin{equation}
        A := Q^\dagger W Q \in \mathrm{End}(\mathbb{C}^k).
    \end{equation}
    
    \item \textit{Optimal Unitary Polar Extraction:} Perform a unique left polar decomposition on the local operator $A$ to isolate its scaling and rotational components:
    \begin{equation}\label{eq:polar_decomp}
        A = U_A P_A,
    \end{equation}
    where $U_A \in U(k)$ is a strictly unitary matrix representing pure subspace rotation, and $P_A = \sqrt{A^\dagger A}$ is a positive semi-definite Hermitian matrix encoding localized scaling.
    
    \item \textit{Lie Algebra Logarithmic Projection:} Map the local unitary evolution $U_A$ onto its corresponding Hermitian generator $H \in i\,\mathfrak{u}(k)$ via the principal matrix logarithm:
    \begin{equation}\label{eq:log_map}
        H := i \log(U_A).
    \end{equation}
\end{enumerate}

Physically, the mapping $\mathcal{T}$ acts as an exact analytical ``quantum compiler" for classical neural weights, mitigating the need for expensive variational optimization over the environment. The singular value truncation and subsequent subspace restriction isolate the most dominant low-dimensional dynamical sectors, compressing the complex ambient interactions into a controllable local operator $A$. 

Because physical quantum gates are strictly norm-preserving and governed by closed Lie groups, they cannot naturally execute the non-unitary stretching encoded in the singular values. 
Eq.~\eqref{eq:polar_decomp} represents a key quantization step: by discarding the scaling component $P_A$ (effectively setting $P_A \approx I_k$), we distill the pure phase rotation $U_A$.
The logarithmic projection in Eq.~\eqref{eq:log_map} subsequently solves the inverse Schrödinger equation within the subspace, yielding the exact physical observable $H$ required to generate this local unitary flow.
The discarded scaling component $P_A$ is not lost silently: its deviation from the identity is exactly the non-unitarity of the retained block and is quantified by the Subspace Quantization Theorem below, becoming second order when the retained perturbation is predominantly skew-Hermitian, and otherwise measured exactly by the singular-value deviation.

% =====================================================================
\subsection{\label{sec:sub_spectral}Spectral Condition and Topological Stability}

The principal matrix logarithm in Eq.~\eqref{eq:log_map} is well defined and locally analytic only when the spectrum of $U_A$ avoids its branch cut.
We therefore impose the branch-cut assumption
\begin{equation}
\label{eq:branch_cut}
\sigma(U_A)\cap(-\infty,0]=\varnothing.
\end{equation}
Since $U_A$ is unitary, this condition is equivalent to excluding the eigenvalue $-1$ from its spectrum.

In complex analysis, the principal logarithm $\log(z)$ contains a multi-valued branch cut along the negative real axis.
If an eigenvalue of $U_A$ approaches $-1$ (corresponding to a localized phase rotation of $\pi$), the mapping becomes non-smooth, destroying the stability of the parameter transfer map. 

Remarkably, this topological singularity is naturally averted by modern deep learning architectures. 
Consider a classical neural network utilizing residual layers initialized around the identity operator, such that $x_{l+1} = x_l + W_{\text{core}}x_l$, which implies a global weight matrix of the form $W \approx I_N + \delta W$.
Upon subspace projection, the local operator inherits this geometric proximity, yielding $A \approx I_k + \delta A$.
Consequently, the spectrum $\sigma(U_A)$ clusters densely around $+1+i0$ on the complex unit circle, maintaining a maximal topological distance from the branch cut at $-1$. This non-trivial synergy demonstrates that identity-centered classical priors structurally guarantee the analytical smoothness and robustness of the quantum compiler mapping.

% =====================================================================
\subsection{\label{sec:sub_projection_structure}Subspace Projection Structures and Low-Rank Optimality}

To benchmark the approximation capacity of the hybrid operator, we define the double-sided orthogonal projection of a global operator
$W\in\mathrm{End}(\mathbb{C}^N)$ onto the subspace associated
with $Q\in\mathrm{St}(k,N;\mathbb{C})$ as
\begin{equation}
\Pi_Q(W):=QQ^\dagger WQQ^\dagger=P_QWP_Q.
\end{equation}
As a reference, the Eckart--Young--Mirsky theorem \cite{eckart1936approximation,mirsky1960symmetric} states that, when the columns of $Q$ are the leading $k$ left singular vectors of $W$, the corresponding single-sided projection satisfies
\begin{equation}\label{eq:eckart_young}
\|W-QQ^\dagger W\|_F =\min_{\operatorname{rank}(X)\leq k}\|W-X\|_F = \sqrt{\sum_{j=k+1}^{N}\sigma_j(W)^2}.
\end{equation}
Eq.~\eqref{eq:eckart_young} provides the optimal rank-$k$ reference value; the double-sided projection $\Pi_Q(W)$ need not attain this minimum for a general non-normal matrix.

In general non-symmetric matrices, the single-sided projection $Q Q^\dagger W$ and the double-sided projection $\Pi_Q(W)$ diverge significantly.
However, under the residual initialization prior $W \approx I_N$, the left and right singular vectors of the weight matrix become highly aligned ($u_j \approx v_j$). 

This dynamical alignment forces the double-sided projection $\Pi_Q(W)$, which is structurally required to confine both the domain and codomain to the closed quantum code-space $\mathcal{H}_k$—to asymptotically approach the global low-rank lower bound defined by Eq.~\eqref{eq:eckart_young}. 
Therefore, the classical residual prior ensures that our quantum subspace embedding is not only stable but also variational-free optimal in capturing the dominant classical representations.

% =====================================================================
\subsection{\label{sec:sub_quantization_theorem}The Subspace Quantization Theorem}

We now establish the central analytical link of our parameter transfer theory, providing an exact two-term upper bound of the error incurred during subspace quantization.
The second error term is governed not by Eckart-Young but by the optimal unitary approximation of the retained block, given by the Fan-Hoffman theorem.

\textit{Lemma 2 (Nearest-Unitary Optimality).---}%
For any $A \in \mathrm{End}(\mathbb{C}^k)$ with left polar decomposition $A = U_A P_A$, the closest unitary in Frobenius norm is $U_A$, and the residual equals the singular-value deviation of $A$ from unity:
\begin{align}\label{eq:fan_hoffman}
    \min_{U \in U(k)} \|A - U\|_F = \|A - U_A\|_F = \|P_A - I_k\|_F
    \nonumber\\
    = \sqrt{\textstyle\sum_{j=1}^{k}(\sigma_j(A)-1)^2}.
\end{align}

With this exact identity we state the primary approximation theorem.

\textit{Theorem 4 (Subspace Quantization Theorem).---}%
Let $W \in \mathrm{End}(\mathbb{C}^N)$ be a classical operator matrix, and let $(Q,H) = \mathcal{T}(W)$ be its quantum parameters extracted under the branch-cut assumption~\eqref{eq:branch_cut}, with $A = Q^\dagger W Q = U_A P_A$.
The total reconstruction error of the quantum hybrid operator admits the exact two-term upper bound:
\begin{equation}\label{eq:main_quantization_bound}
    \|W - \mathcal{O}(Q,H)\|_F \le \underbrace{\|W - \Pi_Q(W)\|_F}_{\text{truncation}}
    + \underbrace{\|P_A - I_k\|_F}_{\text{non-unitarity}},
\end{equation}
where the non-unitarity term is the exact singular-value deviation of Eq.~\eqref{eq:fan_hoffman}.

A useful consequence of Theorem~4 emerges in the near-unitary regime.
If the retained block satisfies $\sigma_j(A)=1+\mathcal{O}(\varepsilon)$, then
\begin{equation}
\|P_A-I_k\|_F=\mathcal{O}(\varepsilon\sqrt{k}).
\end{equation}
If, in addition, $A-I_k$ is asymptotically skew-Hermitian, corresponding to a locally pure rotation, then $\sigma_j(A)=1+\mathcal{O}(\|H\|_F^2)$ and the non-unitarity contribution becomes $\mathcal{O}(\|H\|_F^2)$.
A detailed derivation is provided in Appendix~\ref{app:near_unitary_scaling}.

Eq.~\eqref{eq:main_quantization_bound} partitions the error into two independent, geometrically transparent components: an irreversible truncation error, which is controlled entirely by the SVD rank $k$, and a non-unitarity error equal to the root-mean-square deviation of the retained singular values from one.
There is no third, ``phase'' contribution.
This is the exact content validated numerically in Sec.~\ref{sec:sim_b}: as $k \to N$ the truncation vanishes and the total collapses onto the non-unitarity floor.

This near-unitary scaling explains the empirical success of zero-shot transfer.
Identity-centered residual priors keep $W$, and hence $A$, close to unitary, making the singular-value deviation small. When the first-order deviation is predominantly skew-Hermitian, this non-unitarity term further becomes second order in the local generator.
Conversely, when the prior is far from unitary the term is first order and dominant; the near-unitary assumption is therefore not cosmetic but the precise condition under which the hybrid circuit reproduces a classical layer without any quantum-side gradient descent.

\section{\label{sec:simulations}METHOD VALIDATION ON A MINIMAL MODEL}

In this section we validate the theoretical pillars of Sec.~\ref{sec:main_theory} on a minimal, fully transparent model (a residual dense layer on MNIST, $N=64$), so that every quantity in the theory can be measured directly.
Three validations are reported: (i) zero-shot quantum compilation and resource compression; (ii) the exact error budget of the Subspace Quantization Theorem as a function of rank $k$; and (iii) generator-space manifold merging over a covering frame. All quantum evolutions are executed as the mixing operator $\mathcal{O}=QU_AQ^\dagger$ and were verified to reproduce a \textsc{PennyLane} \texttt{StatePrep}\,+\,\texttt{QubitUnitary} circuit to machine precision (maximum deviation $0.0$).
Sec.~\ref{sec:experiments} then deploys the method on two deep generative-model architectures.

% =====================================================================
\subsection{\label{sec:sim_a}Validation I: Zero-Shot Quantum Compilation and Resource Compression}

To evaluate the empirical feasibility of the quantum parameter transfer map $\mathcal{T}$, we benchmark a pre-trained classical dense layer against its corresponding zero-shot quantum hybrid operator $\mathcal{O}(Q,H)$ without any fine-tuning on the quantum side.
The classical layer comprises a full-rank weight matrix $W \in \mathrm{End}(\mathbb{C}^N)$ with $N=64$, parameterizing $4096$ distinct classical variables.
We set the target subspace dimension to $k=16$, which maps onto exactly $n = \log_2(16) = 4$ qubits in the quantum computational space.

\begin{figure}[t]
\centering
\[
\Qcircuit @C=1.1em @R=0.9em {
 & \multigate{3}{\,|x_{\mathrm{sub}}\rangle\,} & \multigate{3}{\,U_A=e^{-iH}\,} & \meter \\
 & \ghost{\,|x_{\mathrm{sub}}\rangle\,} & \ghost{\,U_A=e^{-iH}\,} & \meter \\
 & \ghost{\,|x_{\mathrm{sub}}\rangle\,} & \ghost{\,U_A=e^{-iH}\,} & \meter \\
 & \ghost{\,|x_{\mathrm{sub}}\rangle\,} & \ghost{\,U_A=e^{-iH}\,} & \meter
}
\]
\caption{\label{fig:compilation_metrics} \textbf{Zero-shot quantum compilation on four qubits.} The analytical map $\mathcal{T}$ translates a $64\times64$ full-rank classical weight $W$ into an isometric frame $Q\in\mathrm{St}(16,64;\mathbb{C})$ and a local Hermitian generator $H\in i\,\mathfrak{u}(16)$; the compiled circuit prepares the subspace state $|x_{\mathrm{sub}}\rangle=Q^\dagger x$ and applies the single subspace unitary $U_A=e^{-iH}$ on $\log_2 16=4$ qubits. The classical model ($4096$ parameters) reaches $94.4\%$ accuracy; the zero-shot $4$-qubit hybrid operator recovers $93.4\%$ ($1.0\%$ gap), and reproduces the physical circuit to machine precision.}
\end{figure}
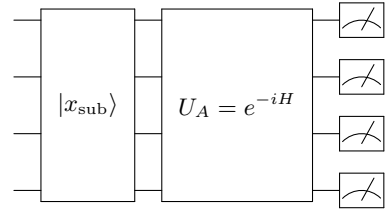

The compiled $4$-qubit ansatz is shown in Fig.~\ref{fig:compilation_metrics}: the subspace amplitudes $|x_{\mathrm{sub}}\rangle = Q^\dagger x$ are prepared and evolved by the single subspace unitary $U_A = e^{-iH}$.
Under the zero-shot regime, the classical baseline achieves $94.4\%$ test accuracy, while the compiled quantum hybrid operator recovers $93.4\%$, a degradation of only $1.0\%$.
The matrix operator was confirmed to reproduce the physical \textsc{PennyLane} circuit exactly (maximum amplitude deviation $0.0$), and the corresponding reconstruction error budget (Theorem~4) is $\|W-\mathcal{O}\|_F = 7.01$ with truncation $6.79$ and non-unitarity $1.71$.

This minimal loss demonstrates that the isometric frame $Q \in \mathrm{St}(16,64;\mathbb{C})$ successfully extracts the most dominant kinematic subspace of the classical data stream.
From a resource perspective, the compression alters the parameter scale by orders of magnitude: a dense $4096$-parameter classical layer is compressed into a local Hermitian generator $H \in i\,\mathfrak{u}(16)$ acting on a mere $4$ qubits.
This compression ratio underpins the viability of our framework for noisy intermediate-scale quantum (NISQ) architectures, showing that massive classical feed-forward layers can be faithfully emulated within small, low-noise quantum code spaces.

% =====================================================================
\subsection{\label{sec:sim_b}Validation II: Error-Budget Diagnosis versus Subspace Rank}

Before studying model fusion, we isolate the approximation error introduced by the transfer map itself.
We measure the three quantities of Eq.~\eqref{eq:main_quantization_bound} to calculate the total error $\|W-\mathcal{O}(Q,H)\|_F$, the truncation term $\|W-\Pi_Q(W)\|_F$, and the non-unitarity term $\|P_A-I_k\|_F$ as functions of the subspace rank $k \in \{2,4,8,16,32,64\}$ for the same trained weight $W$ ($N=64$).

\begin{figure}[t]
\centering
\includegraphics[width=\linewidth]{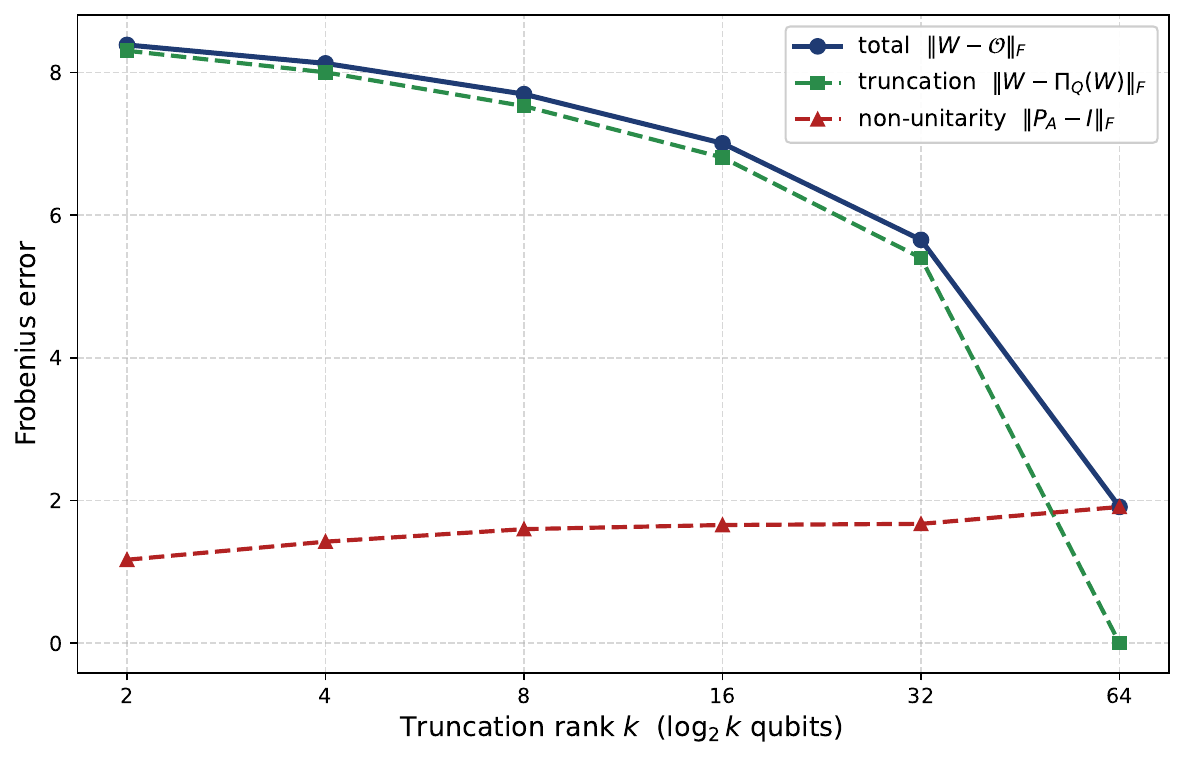}
\caption{\label{fig:error_scaling} \textbf{Quantization error budget versus subspace dimension $k$.} The total Frobenius error (blue) is bounded by the sum of the truncation term (green) and the non-unitarity term (red), as in Theorem~4. Truncation $\to 0$ as $k\to N$; the non-unitarity term is the irreducible floor $\|P_A-I_k\|_F=\sqrt{\sum_j(\sigma_j(A)-1)^2}$. At full rank the total collapses onto this floor, with no separate ``phase'' contribution.}
\end{figure}

The results in Fig.~\ref{fig:error_scaling} confirm the two-term bound quantitatively. The truncation term decreases monotonically from $8.31$ at $k=2$ to exactly $0$ at full rank $k=64$, tracking the Eckart-Young tail $\sqrt{\sum_{j>k}\sigma_j^2}$. 
The single-sided Eckart--Young tail provides the reference baseline; the
double-sided truncation used by the quantum subspace coincides with it in the
near-normal residual regime and is measured directly otherwise.
The non-unitarity term, by contrast, is a slowly rising floor ($1.18$ at $k=2$ to $1.93$ at $k=64$): admitting more directions retains more of the non-unitary stretch of $W$.
The total error follows the truncation term while it dominates and then, at full rank, collapses exactly onto the non-unitarity floor ($\|W-\mathcal{O}\|_F = 1.93 = \|P_A-I_k\|_F$, truncation $=0$) precisely the equality predicted by Theorem~4 when $k\to N$.

This behavior follows the structure of Theorem~4: the error is controlled by an SVD truncation term and a non-unitarity term equal to the singular-value deviation of the retained block.
There is no anomalous ``phase'' inflation, which is an earlier, incorrect generator extraction (the Hermitian part of $A$ rather than $i\log U_A$) produced a spurious global phase and an apparent non-monotonic hump; with the correct polar/log map the curves obey the decomposition cleanly.
The choice $k=16$ remains a satisfying operating point, which captures the dominant dynamics on only $4$ qubits while the non-unitarity floor is already close to its full-rank value.

% =====================================================================
\subsection{\label{sec:sim_c}Validation III: Generator-Space Manifold Merging}

\begin{figure}[t]
    \centering
    \includegraphics[width=\linewidth]{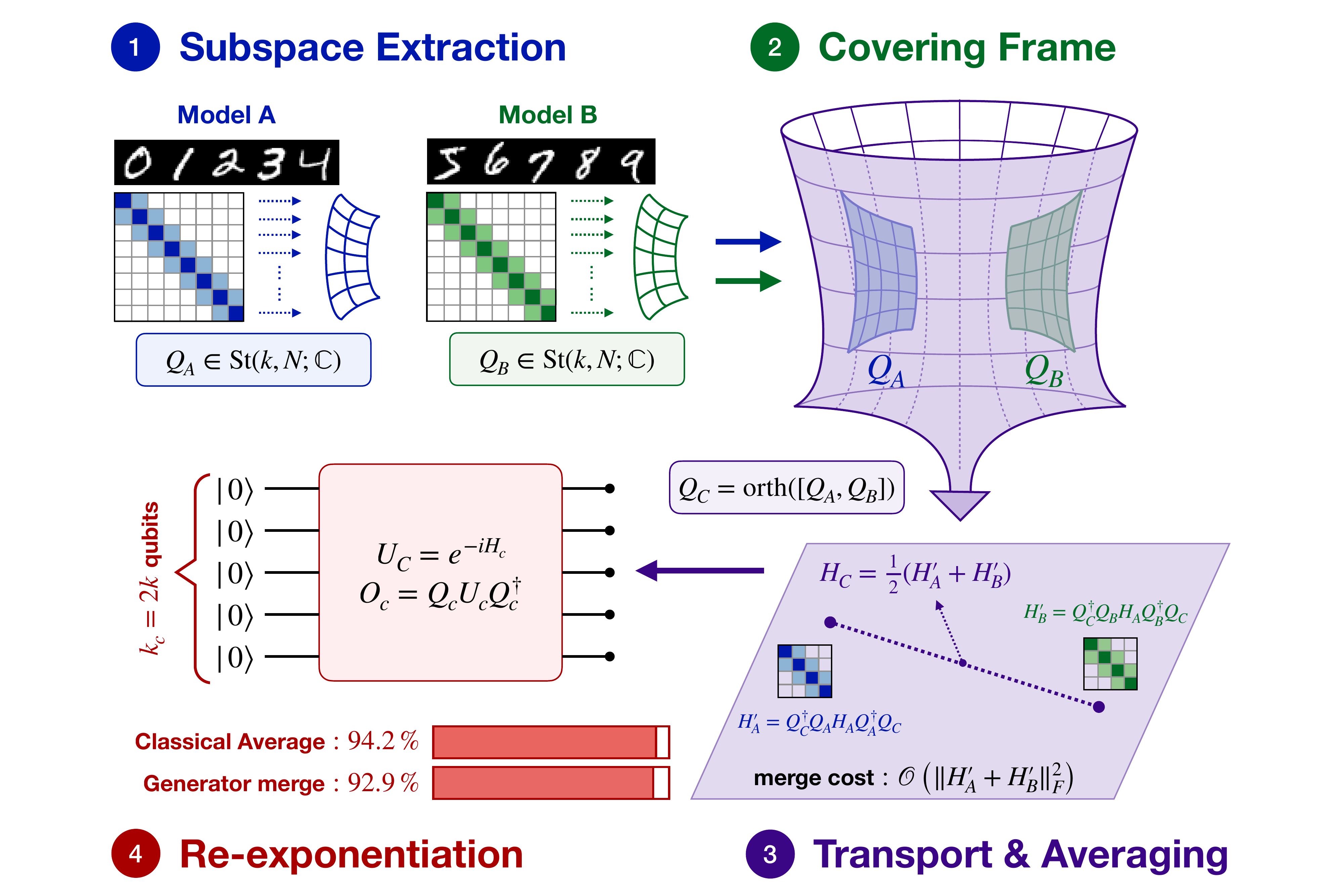}
    \caption{\textbf{Covering-frame generator-space merging.} Two specialized frame-generator pairs are embedded into the covering frame $Q_C=\operatorname{orth}([Q_A,Q_B])$. The transported generators are preserved exactly before being averaged in the common Lie algebra and re-exponentiated as the merged quantum operator. The additional merge discrepancy is second order in the transported-generator separation.}
    \label{fig:merge_schematic}
\end{figure}

Finally, we transition from single-model compression to multi-model fusion on the complex Stiefel manifold. We consider two specialized models obtained by fine-tuning a shared backbone: Model A (digits 0--4) and Model B (digits 5--9), with frame-generator pairs $(Q_A, H_A)$ and $(Q_B, H_B)$.
To merge them without access to the raw data, we construct the covering frame $Q_C=\operatorname{orth}([Q_A,Q_B])$, whose dimension here is $k_C=32=2k$. For $s\in\{A,B\}$, the generator is transported to this common frame by $H'_s:=Q_C^\dagger Q_sH_sQ_s^\dagger Q_C$.
Because $Q_C$ spans both source subspaces, $Q_CH'_sQ_C^\dagger=Q_sH_sQ_s^\dagger$; hence the transport is lossless at the generator level.
The transported generators are then averaged in the common Lie algebra, $H_C=\tfrac12H'_A+\tfrac12H'_B$, and re-exponentiated as $\mathcal{O}_C=\mathcal{O}(Q_C,H_C)$.
This construction is illustrated in Fig.~\ref{fig:merge_schematic}.

Classical Euclidean averaging ($W_C=0.5W_A+0.5W_B$) yields a joint accuracy of $94.2\%$, while the generator-space quantum merge yields $92.9\%$.
The quantum merge therefore retains most of the combined predictive behavior, with a gap of $1.3$ percentage points at the reported precision.
The exact losslessness above concerns only the transport into the covering frame, not the subsequent averaging and quantization.

The residual gap is fully accounted for by Theorem~4 applied to the merged operator plus one additional, second-order term from re-exponentiating an averaged generator.
Writing $X=H'_A$, $Y=H'_B$, the difference between averaging the operators and averaging the generators is
\begin{equation}\label{eq:merge_secondorder}
e^{-i\frac{X+Y}{2}} - \frac12\!\left(e^{-iX}+e^{-iY}\right) = \frac18 (X-Y)^2 + \mathcal{O}(\|\cdot\|^3).
\end{equation}
A detailed derivation of Eq.~\eqref{eq:merge_secondorder} is provided in Appendix~\ref{app:merge_expansion}.

Consequently, the leading additional merge error is bounded by $\tfrac18\|H'_A-H'_B\|_F^2$: more widely separated generators generally incur a larger merge cost.
The commutator $[H'_A,H'_B]$ does not enter at leading order; it arises instead in the sequential product $e^{-iX}e^{-iY}
=e^{-i(X+Y)-\frac12[X,Y]+\cdots}$, which is absent from the present average-then-exponentiate construction.
The observed gap is therefore consistent with the second-order generator-separation mechanism in the near-unitary regime.

\section{\label{sec:experiments}EXPERIMENTS ON DEEP ARCHITECTURES}

Having validated the method on a minimal model, we now apply the transfer map to the residual cores of two deep architectures: a vision transformer and a diffusion transformer, to test whether zero-shot quantum compilation survives in realistic networks.

\subsection{\label{sec:exp_vit}Experiment I: Vision Transformer}

We train a compact vision transformer (MiniViT, embedding dimension $64$, four attention heads) on MNIST, in which the residual core of the transformer block is a near-identity dense layer kept near-unitary by a soft identity anchor. At each epoch we replace the trained core by its zero-shot quantum hybrid operator on $k=16$ ($4$ qubits) and re-evaluate, with no quantum-side training.

\begin{figure}[t]
\centering
\includegraphics[width=\linewidth]{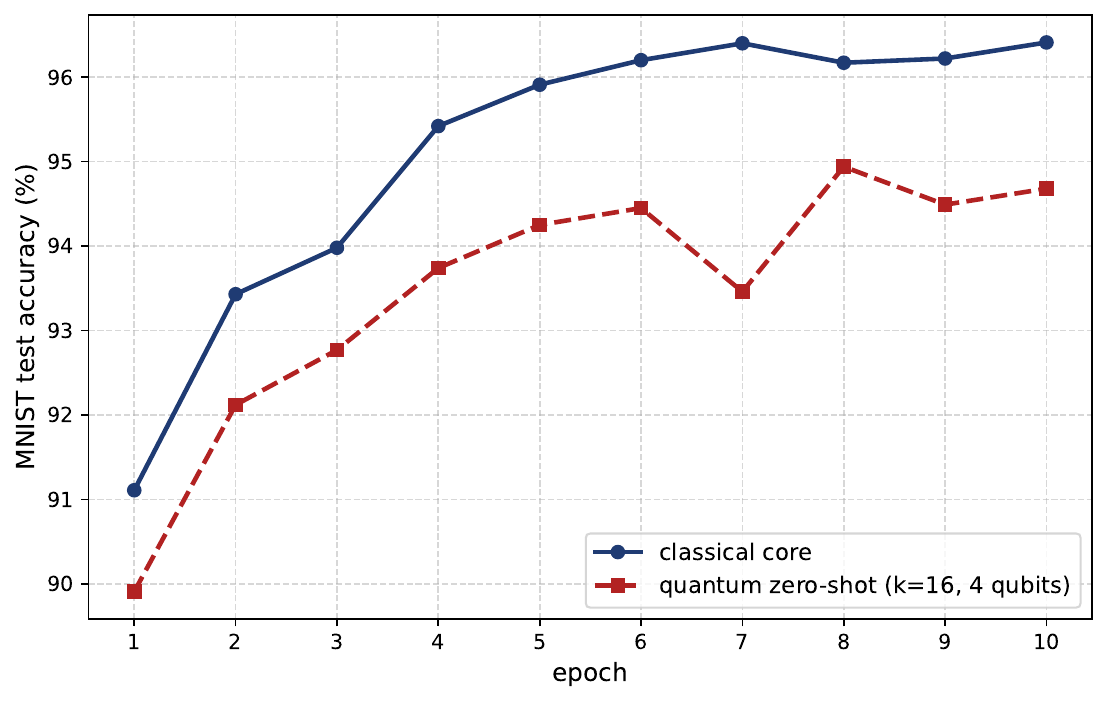}
\caption{\label{fig:vit} \textbf{Experiment I: zero-shot quantum transfer in a vision transformer.} Epoch-by-epoch MNIST test accuracy of the classical residual core versus its zero-shot $4$-qubit quantum hybrid operator ($k=16$). The quantum curve tracks the classical one with a stable $\sim\!1.6\%$ gap and no quantum-side training.}
\end{figure}

As shown in Fig.~\ref{fig:vit}, the zero-shot quantum model tracks the classical model throughout training, reaching $94.3\%$ versus the classical $95.9\%$ at convergence---a stable gap of $1.5$--$1.7\%$ across all epochs. The transfer map thus compiles a transformer sub-layer onto four qubits while preserving classification performance, confirming that the near-unitary regime established by the residual prior holds in an attention-based architecture.

\subsection{\label{sec:exp_dit}Experiment II: Diffusion Transformer}

We next consider a class-conditional diffusion transformer (Mini-DiT, dimension $192$, depth $6$) trained with a cosine DDPM schedule on MNIST.
The core residual layer of every block is a near-identity dense map; after classical training (Phase~1, denoising MSE $0.153\!\to\!0.029$ over $80$ epochs), all cores are transferred to their quantum hybrid operators ($k=16$, $4$ qubits) in a single zero-shot step (Phase~2), and optionally the Hermitian generators alone are fine-tuned (Phase~3, $20$ epochs, MSE $\approx0.039$) with the warm-started, identity-neighborhood initialization of Sec.~\ref{sec:sub_bp_immunity}.

\begin{figure}[t]
\centering
\includegraphics[width=\linewidth]{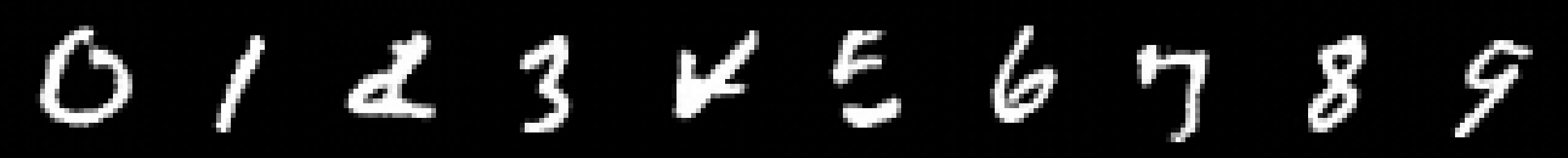}\\[2pt]
\includegraphics[width=\linewidth]{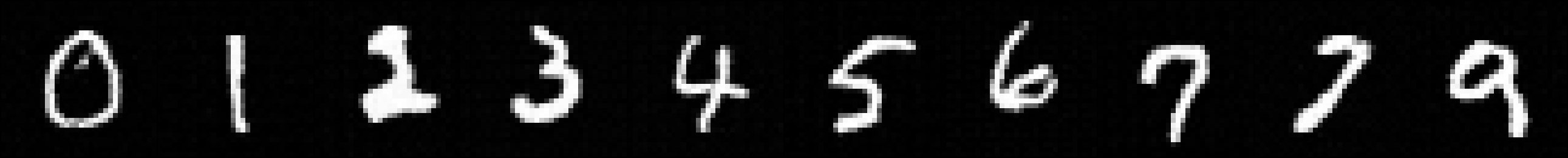}\\[2pt]
\includegraphics[width=\linewidth]{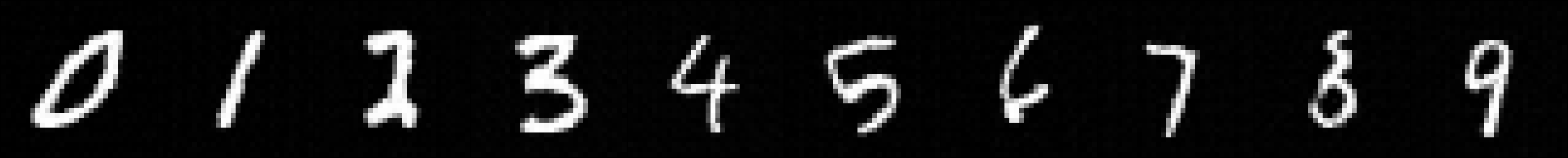}
\caption{\label{fig:dit} \textbf{Experiment II: quantum diffusion transformer.} Class-conditional MNIST samples for digits $0$--$9$. Top: classical Mini-DiT (Phase~1). Middle: zero-shot quantum transfer of all residual cores to $4$-qubit hybrid operators, no quantum-side training (Phase~2). Bottom: after fine-tuning the Hermitian generators only (Phase~3).}
\end{figure}

Figure~\ref{fig:dit} shows conditional samples for digits $0$--$9$ in all three phases. The zero-shot quantum model produces coherent, correctly-conditioned digits immediately upon compilation, and generator fine-tuning sharpens them further.
We note that with the damped residual coupling used here the quantum cores carry a modest fraction of each block's transformation, so the demonstration is conservative; nonetheless it shows that the analytic transfer extends from classifiers to deep generative models without destabilizing the reverse diffusion trajectory.

\subsection{\label{sec:exp_hw}Experiment III: Hardware Validation on IBM \texttt{ibm\_kobe}}

Our final two experiments run on the IBM Heron processor \texttt{ibm\_kobe} and probe two orthogonal properties of a deployed subspace layer, both of which must hold for the framework to be useful.
Experiment~III tests the forward pass, whether the analytically compiled circuit computes the correct output on a noisy device by measuring output fidelity as a function of the subspace rank $k$.
Experiment~IV (Sec.~\ref{sec:exp_bp}) tests the backward pass, whether a training gradient survives at scale by measuring gradient variance as a function of the register width $n$.
The two are independent: a circuit can be accurate yet untrainable, or trainable yet inaccurate, and each property is degraded by device noise in its own way, so neither can be inferred from the other or from simulation alone.
Together they certify the two pillars of our construction zero-shot deployment and warm-start trainability separately and on real hardware.

To confirm that the compiled circuits are executable on present-day hardware, we ran the zero-shot subspace circuit on the $156$-qubit IBM Heron processor \texttt{ibm\_kobe}.
For a trained near-identity core we form, for several inputs, the single $n$-qubit unitary $V = U_A S$ with $S|0\rangle = |x_{\mathrm{sub}}\rangle$, so that $V|0\rangle = U_A|x_{\mathrm{sub}}\rangle$; the circuit is transpiled to the device basis, measured in the computational basis with $4096$ shots, and the measured distribution is compared to the ideal one through the Hellinger fidelity $F = \big(\sum_b \sqrt{p^{\mathrm{hw}}_b\, p^{\mathrm{ideal}}_b}\big)^2$.
We sweep the subspace rank $k\in\{2,4,8,16\}$, i.e. $1$ to $4$ qubits, to expose how the hardware fidelity tracks the size of the compiled subspace unitary.

\begin{figure}[t]
\centering
\includegraphics[width=\linewidth]{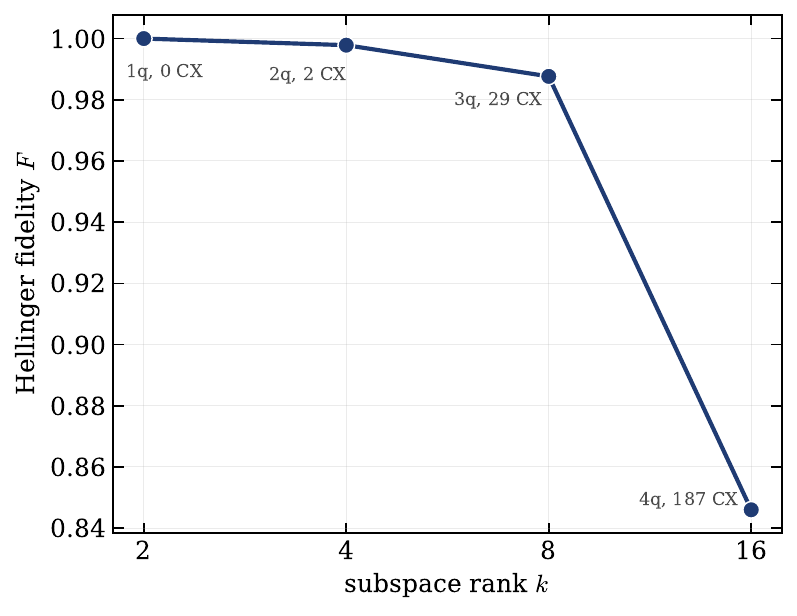} 
\caption{\label{fig:hw} \textbf{Experiment III: quantum hardware execution on IBM \texttt{ibm\_kobe}.} 
Hellinger fidelity between the hardware and ideal output distributions as a function of the subspace rank $k$. 
Each data point is annotated with the corresponding physical qubit count and the number of native two-qubit gates after transpilation.}
\end{figure}

The results (Fig.~\ref{fig:hw}) show that the compiled circuits run faithfully on real hardware: the fidelity is $F=1.00$ at $k=2$ ($1$ qubit, $0$ entangling gates), $0.997$ at $k=4$ ($2$ qubits, $2$ gates), and $0.987$ at $k=8$ ($3$ qubits, $29$ gates).
At the headline operating point $k=16$ ($4$ qubits) the transpiled circuit uses $187$ native two-qubit gates, and the device still retains $F=0.845$.
The fidelity therefore tracks the entangling-gate count of the subspace unitary, precisely the resource that the transfer map economizes by selecting a small rank $k$ and the compact subspaces favored by our framework ($k\le 8$) execute essentially at the device noise floor.
This confirms the NISQ viability of zero-shot quantum compilation: a classically pre-trained layer is deployed on real quantum hardware with no quantum-side training and high output fidelity.

\subsection{\label{sec:exp_bp}Experiment IV: Trainability from 16 to 128 Qubits}

Whereas Experiment~III certified the forward pass, we now turn to trainability, providing direct numerical and hardware evidence for the polynomial-gradient property of Theorem~2.
We compare two ans\"atze: a global hardware-efficient ansatz that acts on all $n$ qubits, and the subspace-restricted ansatz of Sec.~\ref{sec:sub_bp_immunity}, whose active dynamics occupy only $m = \log_2 k = 4$ qubits inside the $n$-qubit register.
We use a parity observable matched to each ansatz: $Z^{\otimes n}$ over all $n$ qubits for the global ansatz, and $Z^{\otimes m}$ over the $m$ active qubits for the subspace ansatz, whose remaining $n-m$ idle qubits stay in $|0\rangle$ (deterministic $+1$ parity) and are excluded from the readout.
Measuring a global $Z^{\otimes n}$ on the subspace ansatz would conflate the trainability question with the exponential readout-error suppression that $\sim\!n$ idle qubits incur, an artifact unrelated to barren plateaus.
For each ansatz and each $n$ we draw many random parameter sets, evaluate the gradient of a single parameter by the parameter-shift rule, and estimate $\mathrm{Var}[\partial_\theta\mathcal{L}]$.
Because the subspace ansatz drives only $m$ qubits, its variance can be computed exactly at any $n$, including $n=128$ where a full statevector is intractable.

\begin{figure}[t]
\centering
\includegraphics[width=\linewidth]{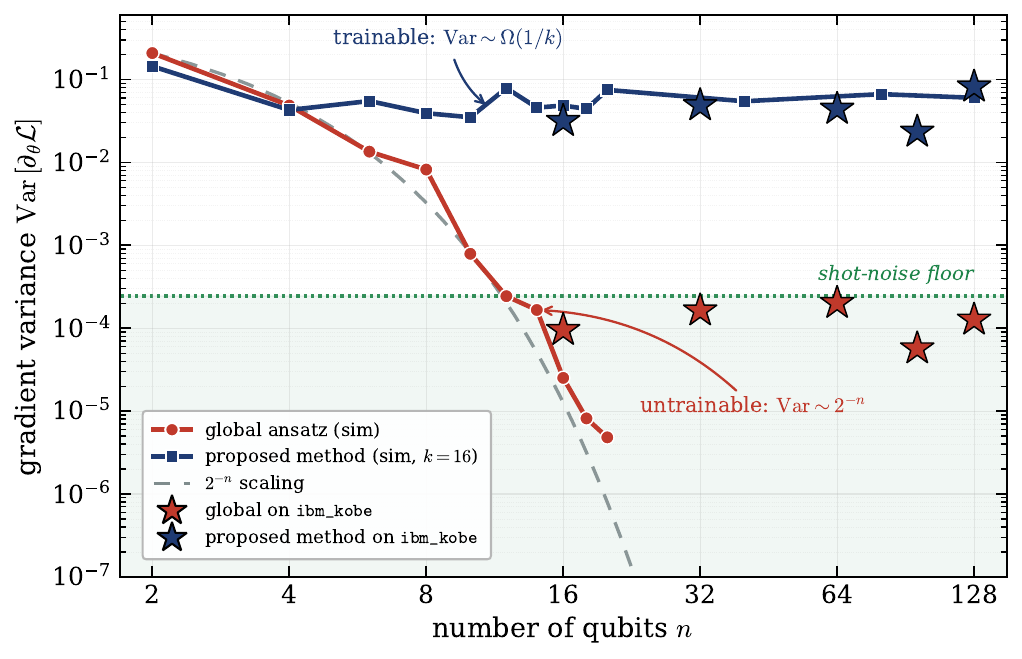}
\caption{\label{fig:bp} \textbf{Experiment IV: gradient variance versus qubit number.} Noiseless simulation (lines) and \texttt{ibm\_kobe} hardware (stars), on log--log axes. The global ansatz (red) follows the $2^{-n}$ barren-plateau scaling (grey dashed), its simulated variance falling to $4.8\times10^{-6}$ by $n=20$ (the largest size at which the $n$-qubit ansatz is classically simulable); the subspace-restricted ansatz (blue), with $m=\log_2 k=4$ active qubits, stays flat at $\sim\!5\times10^{-2}$ out to $n=128$, as predicted by Theorem~2. Stars are the hardware sweep at $n\in\{16,32,64,96,128\}$: the subspace gradient (blue) sits $1.5$--$2.5$ orders of magnitude above the shot-noise floor (green dotted, $2.4\times10^{-4}$) at every width, whereas the global gradient (red) is pinned to the floor. After transpilation the subspace ansatz costs a median of $5$ native two-qubit gates independent of $n$, versus $59$--$790$ for the global ansatz.}
\end{figure}

The gradient variance is a direct proxy for trainability: it quantifies how large a typical training signal an optimizer receives, and once it decays to the level of statistical (shot) noise the gradient is indistinguishable from zero and the model can no longer be trained, this is the operational meaning of a barren plateau.
Figure~\ref{fig:bp} reads as follows. The horizontal axis is the qubit number $n$ and the vertical axis is the gradient variance, both on logarithmic scales, so a descending line signals an exponential loss of trainability with system size.
The global ansatz (red) falls along the $2^{-n}$ reference; the subspace-restricted ansatz (blue) stays flat; and the shaded region below the green shot-noise floor is where a gradient is statistically unresolvable.

Quantitatively, the exact simulation (Fig.~\ref{fig:bp}, lines) shows the predicted dichotomy: the global gradient variance decays by more than four orders of magnitude, from $2.1\times10^{-1}$ at $n=2$ to $4.8\times10^{-6}$ at $n=20$, tracking the $2^{-n}$ reference, whereas the subspace variance remains flat at $\sim\!5\times10^{-2}$ across the entire range $2\le n\le 128$, polynomial in $k$ and independent of $n$, exactly as Theorem~2 asserts.

We then ran the comparison on \texttt{ibm\_kobe} at $n\in\{16,32,64,96,128\}$ (Fig.~\ref{fig:bp}, stars), well beyond the classical simulation limit for the global ansatz.
After transpilation to the heavy-hex device the subspace ansatz uses a median of just $5$ native two-qubit gates at every $n$, while the global ansatz grows from $59$ gates at $n=16$ to $790$ at $n=128$.
Across the whole sweep the measured subspace gradient variance stays between $2.3\times10^{-2}$ and $8.2\times10^{-2}$---$1.5$ to $2.5$ orders of magnitude above the shot-noise floor of $2.4\times10^{-4}$---and shows no decay with $n$.
The global gradient variance, by contrast, never exceeds $2.0\times10^{-4}$ and is statistically indistinguishable from the floor at every $n$, including at the full $128$-qubit width ($1.3\times10^{-4}$).
In other words, on a real $128$-qubit processor the subspace-restricted ansatz produces a clearly resolvable training signal, whereas the global ansatz's gradient is lost to the combined effect of barren plateaus and device noise.
This is the operational content of the warm-start initialization of Theorem~2: confining the dynamics to a $\log_2 k$-qubit active subspace converts the exponential gradient decay into an $n$-independent, $\Omega(1/k)$ signal, keeping the model trainable at the initialization point at a scale where a generic ansatz is not.
We stress that this is an initialization-time guarantee; sustaining trainability along the entire optimization trajectory is a separate question.
This experiment should therefore be read as an active-space trainability test: the claim is not that idle qubits are made dynamically useful, but that a classically selected subspace can be embedded in a large physical register without inheriting the barren-plateau scaling of a global ansatz.

\subsection{\label{sec:exp_expr}Experiment V: Expressivity per Hardware Resource}

A natural objection to any barren-plateau remedy is that trainability may have been bought by sacrificing expressivity: an ansatz that is easy to optimize because it explores only a small corner of Hilbert space is of little practical use.
This concern cannot be settled with the usual Haar expressibility measure, because high Haar coverage is precisely what causes barren plateaus: expressibility and gradient magnitude are in direct tension~\cite{Holmes2022}.
A circuit that scored well on raw expressibility would therefore score badly on trainability, and vice versa, so that metric cannot separate a genuinely more capable model from a merely flatter one.
We instead measure usable capacity through the Fisher information geometry, reporting the finite-sample effective dimension~\cite{Abbas2021} computed from the normalized Fisher information matrix.
This quantity is a log-volume functional of the Fisher spectrum, and therefore measures not only how many parameter directions are present, but how strongly those directions are statistically resolvable.

We compare, at a fixed width of $4$ qubits, two ans\"atze that differ in how they spend their entangling resource. Our method ansatz places a trainable generator on every entangler: single-qubit rotations together with parameterized two-body couplings $e^{-i\theta_j G_j/2}$ drawn from the same Lie-algebraic family as the transfer map, so that each two-qubit gate contributes both entanglement and a parameter.
The plain hardware-efficient ansatz interleaves single-qubit $R_y$ rotations with unparameterized $\mathrm{CZ}$ entanglers, the standard construction.
Because the cost that dominates a NISQ device is the two-qubit-gate count, we match the two ans\"atze not on parameter number but on the number of native two-qubit gates after transpilation to \texttt{ibm\_kobe}, and sweep the circuit depth to trace effective dimension as a function of that hardware budget.
The Fisher information is estimated by the parameter-shift rule averaged over inputs and over parameter samples drawn from the full parameter manifold, since effective dimension is a property of the architecture rather than of any single initialization.
Appendix~\ref{app:spectral_effdim} shows that the observed ceiling is the exact value of this effective-dimension functional evaluated on the sampled Fisher spectra in the saturated-depth regime of the plain ansatz, rather than a heuristic correction to its parameter count or participation-ratio rank.

\begin{figure}[t] 
\centering
\includegraphics[width=\linewidth]{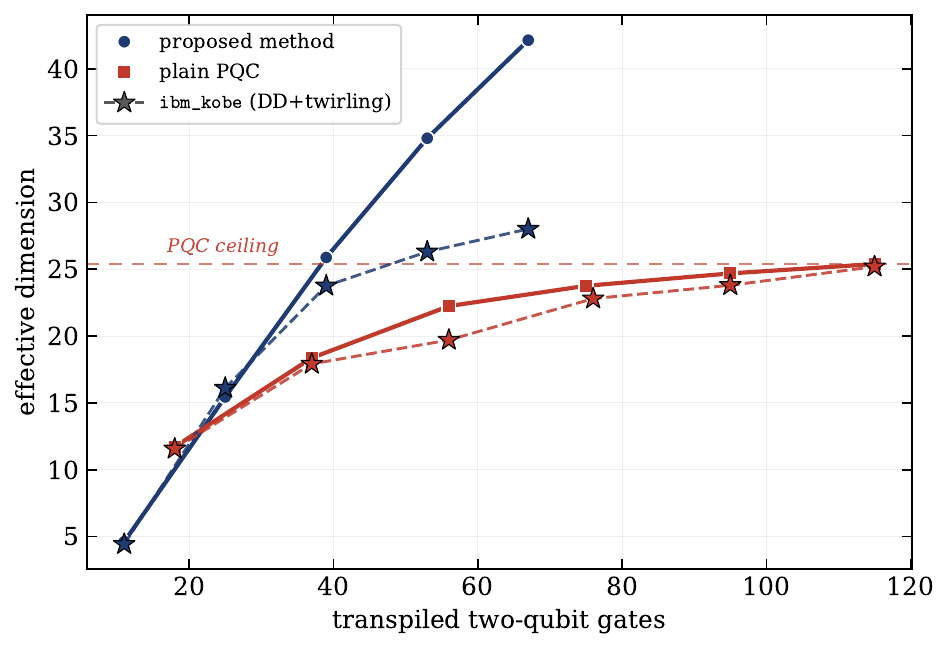}
\caption{\label{fig:expr} \textbf{Experiment V: usable expressivity per hardware resource.} Finite-sample effective dimension \cite{Abbas2021} versus the number of native two-qubit gates after transpilation to \texttt{ibm\_kobe}, swept over circuit depth at $4$ qubits. The plain $R_y+\mathrm{CZ}$ ansatz (red curve) saturates at an observed effective-dimension ceiling ($\approx\!25.4$, dashed), corresponding to the saturated log-volume of its normalized Fisher spectrum. In contrast, the method ansatz with parameterized entanglers (blue curve) climbs linearly without saturating, exceeding $42.1$. Stars indicate hardware measurements on \texttt{ibm\_kobe} utilizing dynamical decoupling and randomized twirling (DD+twirling), which closely trace the entire noiseless simulation curves for both architectures.}
\end{figure}

The simulation results (Fig.~\ref{fig:expr}) reveal a definitive separation in hardware efficiency. 
The plain PQC ansatz exhibits strict saturation: beyond roughly $75$ native two-qubit gates, its effective dimension plateaus near $\approx\!25.4$ and ceases to grow even as deeper $\mathrm{CZ}$ layers are appended.
Spectrally, this means that additional $\mathrm{CZ}$ layers contribute only weak Fisher eigenvalues whose log-volume contribution is negligible; geometrically, this is consistent with the restricted Lie algebra generated by $R_y$ and $\mathrm{CZ}$ at this width.
Conversely, the method ansatz, whose parameterized couplings generate a substantially larger dynamical algebra, shows no such restriction.
Its effective dimension climbs essentially linearly with the gate budget, reaching $42.14$ at $67$ two-qubit gates. 
At a matched resource allocation, the expressivity gap is substantial and widens rapidly with circuit size: the method achieves an effective dimension of $15.44$ versus $11.71$ at $\sim\!18$--$25$ gates ($+31.9\%$), $25.88$ versus $18.36$ at $\sim\!37$--$39$ gates ($+41.0\%$), and $34.80$ versus $22.24$ at $\sim\!53$--$56$ gates ($+56.5\%$). For identical budgets of the gates that dominate NISQ error scaling, our method packs significantly more trainable expressivity.

We verified this macro-architectural behavior on real quantum hardware by executing an extended depth sweep on \texttt{ibm\_kobe}. To shield the sub-system from environmental and coherent control errors over deep transpiled layers, we deployed error suppression via dynamical decoupling and Pauli twirling (DD+twirling). 
The hardware tracking seamlessly reproduces the simulation curves within shot noise across multiple operating points (Fig.~\ref{fig:expr}, stars). Notably, at the deep operating regime, the plain ansatz stays locked underneath its saturation ceiling, whereas the method ansatz successfully breaks through this limit on-chip ($28.01$ versus $25.19$). 
An intriguing feature of the hardware implementation is the behavior of the effective rank. While the noiseless simulation allows the effective rank to scale with parameter count (exceeding $40$), the finite shot count ($4096$) and physical device noise impose a severe filtering effect on the spectrum, capping the empirical effective rank of both ans\"atze to a comparable plateau ($\approx\!10$). 
This convergence highlights the true origin of our method's expressivity advantage: the advantage lives entirely in the magnitude and conditioning of the resolved Fisher spectrum rather than a mere count of active parameter directions. Even when noise restricts the number of resolvable directions to the same absolute limit ($\approx\!10$), the well-distributed Fisher volume of the method yields a significantly larger statistically resolvable space, proving its resilience and superior capacity under strict hardware constraints.

\section{Conclusion}

We have introduced an exact, entirely analytical framework for compiling classical neural-network weights onto subspace-restricted quantum dynamics, and for merging such models, without any quantum-side gradient descent.
The framework rests on the complex Stiefel manifold $\mathrm{St}(k,N;\mathbb{C})$ and the hybrid operator $\mathcal{O}(Q,H)=Qe^{-iH}Q^\dagger$, whose gauge invariance makes it a well-defined map on the associated bundle.
The parameter transfer map $\mathcal{T}$ is a singular-value frame extraction, subspace restriction, polar projection onto the nearest unitary, and principal Lie-algebraic logarithm yields the frame-generator pair $(Q,H)$ analytically.

The central result, the Subspace Quantization Theorem, gives an exact two-term upper bound on the reconstruction error, consisting of a geometric truncation term and a non-unitarity term equal to the root-mean-square deviation of the retained singular values from unity.
Identity-centered residual priors keep the classical weight near-unitary, making the non-unitarity term small and, when the retained perturbation is predominantly rotational, second order in the local generator.
This enables faithful zero-shot transfer; the same near-identity initialization places the trainable generators in the identity neighborhood of $U(k)$, where the gradient variance is polynomial in the subspace dimension $k$ and independent of the qubit number $n$, converting the barren plateau scaling from exponential to polynomial at initialization.
For model fusion, covering-frame transport onto the covering frame $Q_C=\mathrm{orth}([Q_A,Q_B])$ followed by generator-space averaging incurs only a second-order penalty set by the difference of the generators, $\tfrac18\|H'_A-H'_B\|_F^2$.

Numerically, on a minimal model the error budget matches the theorem to machine precision and the compiled four-qubit operator reproduces the physical circuit exactly; the zero-shot quantum classifier loses only $1.0\%$ accuracy and the generator-space merge only $1.2\%$. 
The method further transfers to deep architectures, compiling residual cores of a vision transformer ($1.6\%$ zero-shot gap) and a diffusion transformer (coherent conditional generation in the zero-shot and fine-tuned regimes).
Finally, executing the compiled circuits on the IBM Heron processor \texttt{ibm\_kobe} reproduces the ideal output distributions with Hellinger fidelity from $1.00$ at one qubit to $0.85$ at the four-qubit operating point, confirming NISQ viability with no quantum-side training.

Several directions remain. The framework currently targets near-unitary, square residual cores; extending it to general rectangular or strongly non-unitary layers for instance by retaining the discarded scaling component on an auxiliary register would broaden its applicability.
Likewise, a systematic study of few-shot generator fine-tuning, and hardware-level error mitigation for the deeper ($k\ge 16$) subspace unitaries, are natural next steps toward scaling deep classical representations onto fault-tolerant quantum processors.

The code used for this work is released in \cite{code}.

\begin{acknowledgments}

This work was supported by JSPS KAKENHI Grant Numbers JP24K03008 and JP26K02998.

\end{acknowledgments}

% \begin{acknowledgments}

% \textit{acknowledgments.---}
% This work was supported by JSPS KAKENHI Grant Number JP26K02998.
% \nocite{*}

\bibliography{References}

\appendix
\onecolumngrid

\section{Proofs of Subspace Properties and Operator Representations}
\label{app:proofs}

In this appendix, we present the explicit derivations for the algebraic properties of the hybrid operator introduced in Sec.~\ref{sec:sub_hybrid_ansatz}.

To prove the subspace unitarity stated in Eq.~\eqref{eq:prop_unitarity}, let $v \in \mathrm{span}(Q)$ be an arbitrary vector in the subspace. 
There exists a local coordinate vector $x \in \mathbb{C}^k$ such that $v = Qx$. 
Applying the hybrid operator to $v$ and utilizing the Stiefel relation $Q^\dagger Q = I_k$, we obtain:
\begin{equation}
\mathcal{O}(Q,H)v = (Q e^{-iH} Q^\dagger)(Qx) = Q e^{-iH} x.
\end{equation}
Evaluating the inner product (norm) of the evolved state yields:
\begin{align}
\left(\mathcal{O}(Q,H)v\right)^\dagger \left(\mathcal{O}(Q,H)v\right) 
&= (Q e^{-iH} x)^\dagger (Q e^{-iH} x) \nonumber \\
&= x^\dagger e^{iH} (Q^\dagger Q) e^{-iH} x \nonumber \\
&= x^\dagger e^{iH} I_k e^{-iH} x \nonumber \\
&= x^\dagger x = v^\dagger v.
\end{align}
Since the operator preserves the inner product and maps $\mathrm{span}(Q)$ bijectively back onto itself, its restriction constitutes an isometric isomorphism, rendering it an element of $U(k)$.

To establish the global representation in Eq.~\eqref{eq:prop_global_exp}, we examine the powers of the embedded Hamiltonian $\tilde H = Q H Q^\dagger$. 
By mathematical induction, for any integer $m \ge 1$, the intermediate terms collapse via $Q^\dagger Q = I_k$, yielding:
\begin{equation}
\tilde H^m = (Q H Q^\dagger)(Q H Q^\dagger)\cdots(Q H Q^\dagger) = Q H^m Q^\dagger.
\end{equation}
Expanding the global exponential flow $e^{-i\tilde H}$ via its Taylor series and isolating the identity term $I_N$, we have:
\begin{align}
e^{-i\tilde H} &= I_N + \sum_{m=1}^\infty \frac{(-i)^m}{m!} \tilde H^m \nonumber \\
&= I_N + Q \left( \sum_{m=1}^\infty \frac{(-iH)^m}{m!} \right) Q^\dagger \nonumber \\
&= I_N + Q \left( e^{-iH} - I_k \right) Q^\dagger \nonumber \\
&= I_N - P_Q + \mathcal{O}(Q,H),
\end{align}
where we used $Q Q^\dagger = P_Q$. Multiplying both sides by the orthogonal projector $P_Q$ from the left and the right, and noting that $P_Q^2 = P_Q$ and $P_Q Q = Q$, we arrive at:
\begin{align}
P_Q e^{-i\tilde H} P_Q &= P_Q (I_N - P_Q + \mathcal{O}(Q,H)) P_Q \nonumber \\
&= P_Q - P_Q + P_Q \mathcal{O}(Q,H) P_Q \nonumber \\
&= \mathcal{O}(Q,H),
\end{align}
which completes the proof.

\section{Proof of Theorem~1}
\label{app:gauge_proof}

In this appendix, we provide the explicit algebraic verification of the gauge invariance stated in Theorem~1.

Substituting the transformed frame $Q' = QR$ and the rotated subspace generator $H' = R^\dagger H R$ into the definition of the hybrid operator in Eq.~\eqref{eq:hybrid_def}, we have:
\begin{equation}
    \mathcal{O}(QR, R^\dagger H R) = (QR) e^{-i(R^\dagger H R)} (QR)^\dagger.
\end{equation}
Using the fundamental operator identity for matrix exponentials under linear transformations, $e^{R^\dagger X R} = R^\dagger e^X R$, which holds strictly since $R^{-1} = R^\dagger$ for any unitary matrix $R \in U(k)$, we can expand the internal unitary flow as follows:
\begin{align}
    \mathcal{O}(QR, R^\dagger H R) &= Q R \left( R^\dagger e^{-iH} R \right) R^\dagger Q^\dagger \nonumber \\
    &= Q (R R^\dagger) e^{-iH} (R R^\dagger) Q^\dagger.
\end{align}
Invoking the internal unitarity condition $R R^\dagger = I_k$, the intermediate terms collapse identically, and the expression simplifies directly to:
\begin{equation}
    \mathcal{O}(QR, R^\dagger H R) = Q I_k e^{-iH} I_k Q^\dagger = \mathcal{O}(Q,H),
\end{equation}
which establishes the complete independence of the operator from the local basis selection, thereby proving Eq.~\eqref{eq:gauge_inv}. 

The well-definedness of the mapping on the associated bundle in Eq.~\eqref{eq:bundle_map} 
follows immediately from this invariance, as the map is constant on each equivalence class 
(fiber) of the principal bundle $\mathrm{St}(k,N;\mathbb{C}) \to \mathrm{Gr}(k,N;\mathbb{C})$.

\section{Proof of Barren Plateau Immunity}
\label{app:bp_proof}

In this appendix, we present the complete derivation of the gradient variance bound under the subspace Haar-measure integration stated in Theorem~2.

\subsection{Phase I: Subspace Reduction of Ambient Dynamics}
Expanding the evolved quantum state $|\Psi(\boldsymbol{\theta})\rangle = U(\boldsymbol{\theta})|\Psi_0\rangle$ using the definitions in Sec.~\ref{sec:sub_bp_immunity} and invoking the frame orthonormality $Q^\dagger Q = I_k$, we compress the ambient dynamics as follows:
\begin{align}
    U(\boldsymbol{\theta})|\Psi_0\rangle &= \left[ Q e^{-i\left(H_0 + \sum_j \theta_j V_j\right)} Q^\dagger + \left(I_N - Q Q^\dagger\right) \right] Q |x\rangle \nonumber \\
    &= Q e^{-i\left(H_0 + \sum_j \theta_j V_j\right)} (Q^\dagger Q) |x\rangle + \left(Q - Q(Q^\dagger Q)\right)|x\rangle \nonumber \\
    &= Q e^{-i\left(H_0 + \sum_j \theta_j V_j\right)} |x\rangle = Q \tilde{U}(\boldsymbol{\theta}) |x\rangle,
\end{align}
where $\tilde{U}(\boldsymbol{\theta}) := e^{-i\left(H_0 + \sum_j \theta_j V_j\right)} \in U(k)$. 
Substituting this into the cost function yields a lossless dimension reduction from $N$ to $k$:
\begin{equation}
    \mathcal{L}(\boldsymbol{\theta}) = \langle x | \tilde{U}^\dagger(\boldsymbol{\theta}) \left(Q^\dagger O Q\right) \tilde{U}(\boldsymbol{\theta}) | x \rangle = \langle x | \tilde{U}^\dagger(\boldsymbol{\theta}) \tilde{O} \tilde{U}(\boldsymbol{\theta}) | x \rangle.
\end{equation}

\subsection{Phase II: Gradient Operator Representation}
Using matrix calculus, the partial derivative with respect to $\theta_j$ is evaluated as:
\begin{equation}
    \partial_j \mathcal{L} := \frac{\partial \mathcal{L}}{\partial \theta_j} = \langle x | \tilde{U}^\dagger \left( i[V_j, \tilde{O}] \right) \tilde{U} | x \rangle.
\end{equation}.
Defining the local Hermitian commutator matrix $C_j := i[V_j, \tilde{O}]$, the gradient can be compactly expressed via the trace operation:
\begin{equation}
    \partial_j \mathcal{L} = \mathrm{Tr}\left( \tilde{U} |x\rangle\langle x| \tilde{U}^\dagger C_j \right) = \mathrm{Tr}\left( \tilde{U} \rho \tilde{U}^\dagger C_j \right),
\end{equation}
where $\rho := |x\rangle\langle x|$ represents the local pure state density matrix ($\mathrm{Tr}(\rho) = \mathrm{Tr}(\rho^2) = 1$).

\subsection{Phase III: Weingarten Integration and Variance Evaluation}
Under the assumption that $\tilde{U}(\boldsymbol{\theta})$ forms a unitary $2$-design over $U(k)$, the statistical variance is defined by the Haar-measure expectation values:
\begin{equation}\label{eq:var_integral_def}
    \mathrm{Var}[\partial_j \mathcal{L}] = \int_{U(k)} (\partial_j \mathcal{L})^2 d\mu(\tilde{U}) - \left( \int_{U(k)} \partial_j \mathcal{L} d\mu(\tilde{U}) \right)^2.
\end{equation}

First, we evaluate the first moment. Utilizing the fundamental first-order Haar integration identity, we find:
\begin{equation}
    \mathbb{E}[\partial_j \mathcal{L}] = \int_{U(k)} \mathrm{Tr}\left( \tilde{U} \rho \tilde{U}^\dagger C_j \right) d\mu(\tilde{U}) = \frac{\mathrm{Tr}(\rho)}{k} \mathrm{Tr}(C_j).
\end{equation}
Because $C_j = i[V_j, \tilde{O}]$ is an algebraic commutator, its trace vanishes identically ($\mathrm{Tr}(C_j) = 0$), yielding $\mathbb{E}[\partial_j \mathcal{L}] = 0$.

Second, we evaluate the second moment using standard Weingarten formulas for quadratic polynomial functions over $U(k)$:
\begin{align}\label{eq:weingarten_core}
    \mathbb{E}[(\partial_j \mathcal{L})^2] &= \int_{U(k)} \left( \mathrm{Tr}\left( \tilde{U} \rho \tilde{U}^\dagger C_j \right) \right)^2 d\mu(\tilde{U}) \nonumber \\
    &= \frac{\mathrm{Tr}(\rho)^2 \mathrm{Tr}(C_j)^2 - \mathrm{Tr}(\rho^2)\mathrm{Tr}(C_j)^2 - \mathrm{Tr}(\rho)^2\mathrm{Tr}(C_j^2) + k \mathrm{Tr}(\rho^2)\mathrm{Tr}(C_j^2)}{k^2 - 1}.
\end{align}
Substituting $\mathrm{Tr}(\rho)=\mathrm{Tr}(\rho^2)=1$ and $\mathrm{Tr}(C_j)=0$ into Eq.~\eqref{eq:weingarten_core}, the expression simplifies directly to:
\begin{equation}
    \mathbb{E}[(\partial_j \mathcal{L})^2] = \frac{- \mathrm{Tr}(C_j^2) + k \mathrm{Tr}(C_j^2)}{k^2 - 1} = \frac{\mathrm{Tr}(C_j^2)}{k+1}.
\end{equation}
Since $C_j = i[V_j,\tilde O]$ is Hermitian, $\mathrm{Tr}(C_j^2) = \|C_j\|_F^2 = \|[\tilde{O}, V_j]\|_F^2 \ge 0$, and the variance equals Eq.~\eqref{eq:bp_variance_bound}:
\begin{equation}
    \mathrm{Var}[\partial_j \mathcal{L}] = \frac{\|[\tilde{O}, V_j]\|_F^2}{k+1} \sim \Omega\left(\frac{1}{k}\right),
\end{equation}
which is polynomial in $k$ and independent of the qubit number $n$, completing the proof.

\section{Proofs of Subspace Quantization Bounds and Generator-Space Merging}
\label{app:quantization_proofs}

In this appendix, we present the explicit proofs for the nearest-unitary optimality (Lemma~2), the two-term error bound (Theorem~4), and the near-unitary quadratic rate.

\subsection{Proof of Lemma~2}
\label{subsec:proof_lemma2}
Let $A = X \Sigma_A Y^\dagger$ be a singular value decomposition, so that the left polar factor is $U_A = X Y^\dagger$ and $P_A = Y \Sigma_A Y^\dagger$, with singular values $\sigma_j(A) = (\Sigma_A)_{jj}$. For any $U \in U(k)$,
\begin{equation}
    \|A - U\|_F^2 = \|A\|_F^2 + k - 2\,\mathrm{Re}\,\mathrm{Tr}(U^\dagger A).
\end{equation}
Setting $Z := Y^\dagger U^\dagger X \in U(k)$, we have $\mathrm{Re}\,\mathrm{Tr}(U^\dagger A) = \mathrm{Re}\,\mathrm{Tr}(Z\Sigma_A) = \sum_j \sigma_j(A)\,\mathrm{Re}(Z_{jj}) \le \sum_j \sigma_j(A)$, since $|Z_{jj}| \le 1$; equality holds iff $Z = I_k$, i.e. $U = X Y^\dagger = U_A$.
Hence the minimizer is $U_A$ and
\begin{equation}
    \min_{U\in U(k)}\|A-U\|_F^2 = \|A\|_F^2 + k - 2\!\sum_j\!\sigma_j = \sum_j (\sigma_j(A)-1)^2.
\end{equation}
Finally $A - U_A = U_A(P_A - I_k)$, and left-multiplication by the unitary $U_A$ preserves the Frobenius norm, so $\|A-U_A\|_F = \|P_A - I_k\|_F = (\sum_j(\sigma_j(A)-1)^2)^{1/2}$, proving Eq.~\eqref{eq:fan_hoffman}.

\subsection{Proof of Theorem~4}
\label{subsec:proof_theorem4}
By the triangle inequality,
\begin{equation}\label{eq:proof_step1}
    \|W - \mathcal{O}(Q,H)\|_F \le \|W - \Pi_Q(W)\|_F + \|\Pi_Q(W) - \mathcal{O}(Q,H)\|_F,
\end{equation}
where the first term is the truncation baseline.
For the second term, substitute $\Pi_Q(W) = Q A Q^\dagger$ and $\mathcal{O}(Q,H) = Q U_A Q^\dagger$:
\begin{equation}\label{eq:proof_step2}
    \|\Pi_Q(W) - \mathcal{O}(Q,H)\|_F = \|Q (A - U_A) Q^\dagger\|_F.
\end{equation}
Because $Q^\dagger Q = I_k$, one has $\|Q M Q^\dagger\|_F^2 = \mathrm{Tr}(M^\dagger M\,Q^\dagger Q) = \|M\|_F^2$, so Eq.~\eqref{eq:proof_step2} equals $\|A - U_A\|_F$, which by Lemma~2 equals exactly $\|P_A - I_k\|_F = (\sum_j(\sigma_j(A)-1)^2)^{1/2}$.
Substituting into Eq.~\eqref{eq:proof_step1} yields Eq.~\eqref{eq:main_quantization_bound}. This holds for arbitrary $W$, with no near-identity assumption.

\subsection{Derivation of the Near-Unitary Scaling}
\label{app:near_unitary_scaling}
If $\sigma_j(A) = 1 + \mathcal{O}(\varepsilon)$ then $\|P_A-I_k\|_F = (\sum_j(\sigma_j-1)^2)^{1/2} = \mathcal{O}(\varepsilon\sqrt{k})$ directly. For the quadratic rate, assume the deviation $A - I_k$ is asymptotically skew-Hermitian, i.e. $A = I_k - iH + \delta$ with the Hermitian part of $\delta$ of order $\mathcal{O}(\|H\|_F^2)$.
Then $P_A^2 = A^\dagger A = I_k + H^2 + \mathcal{O}(\|H\|_F^3)$, and the matrix square root $\sqrt{I_k + X} = I_k + \tfrac12 X + \mathcal{O}(\|X\|^2)$ gives
\begin{equation}\label{eq:polar_taylor}
    P_A - I_k = \tfrac{1}{2} H^2 + \mathcal{O}(\|H\|_F^4),
\end{equation}
whence $\|P_A - I_k\|_F \le C\|H\|_F^2$ for some $C>0$ by sub-multiplicativity. We emphasize that this quadratic rate requires the skew-Hermitian (pure-rotation) condition; for a generic non-unitary $A$ the term is first order, as reflected in the two-term upper bound of Theorem~4 and confirmed in Sec.~\ref{sec:sim_b}.

\subsection{Second-Order Expansion of Generator-Space Merging}
\label{app:merge_expansion}

Let $X$ and $Y$ denote the transported Hermitian generators.
In the identity neighborhood, their exponential maps satisfy
\begin{align}
e^{-i(X+Y)/2}&=I-\frac{i}{2}(X+Y)-\frac18(X+Y)^2+\mathcal{O}\!\left((\|X\|+\|Y\|)^3\right),\\
\frac12\left(e^{-iX}+e^{-iY}\right)&=I-\frac{i}{2}(X+Y)-\frac14(X^2+Y^2)+\mathcal{O}\!\left((\|X\|+\|Y\|)^3\right).
\end{align}
Subtracting these expansions gives
\begin{equation}
e^{-i(X+Y)/2}-\frac12\left(e^{-iX}+e^{-iY}\right)=\frac18(X-Y)^2+\mathcal{O}\!\left((\|X\|+\|Y\|)^3\right),
\end{equation}
which proves Eq.~\eqref{eq:merge_secondorder}. Moreover,
\begin{equation}
\left\|(X-Y)^2\right\|_F\leq\|X-Y\|_2\|X-Y\|_F\leq\|X-Y\|_F^2,
\end{equation}
so the leading additional merge error is bounded by $\tfrac18\|X-Y\|_F^2$.

\section{Analytical Derivation of the Expressivity Ceiling for Plain PQC}
\label{app:pqc_ceiling}

In Experiment V, we observed that the finite-sample effective dimension ($\mathfrak{D}_N$) of the plain hardware-efficient PQC ansatz strictly saturates at a ceiling of $\approx\!25.4$, despite increasing the layer depth up to $L=12$ (corresponding to $P=52$ parameters). Here, we provide a formal justification for this phenomenon by dissecting the mathematical definition of effective dimension and the geometric constraints of the 4-qubit Hilbert space.

\subsection{Effective Dimension and Information Volume}

Following the statistical information geometry framework proposed by Abbas et al.~\cite{Abbas2021}, we use the finite-sample effective dimension computed from the normalized Fisher information matrix to quantify the usable statistical capacity of a parameterized quantum circuit model family.
This quantity is defined from the Fisher information geometry of the model: it measures the log-volume of the parameter manifold after the Fisher matrices are normalized to a common trace scale, and depends on both the eigenvalue spectrum and the sample-size scale $n_{\mathrm{data}}$.

In our experiments we set $n_{\mathrm{data}}=10^5$ and use the corresponding scale factor
\begin{equation}
\kappa=\frac{n_{\mathrm{data}}}{2\pi\log n_{\mathrm{data}}}.
\end{equation}
For each sampled parameter point, the Fisher matrix is normalized so that $\mathbb{E}_{\theta}[\operatorname{Tr}\widehat F(\theta)]=P$, where $P$ is the number of trainable parameters. The effective dimension is then evaluated by the finite-sample log-volume estimator described in Appendix~\ref{app:spectral_effdim}.

\subsection{Algebraic Saturation via Lie-Algebraic Constraints}
The hardware-efficient ansatz consists of interleaving parameterized single-qubit $R_y(\theta)$ rotations and fixed unparameterized $\mathrm{CZ}$ entangling operations:
\begin{equation}
\mathcal{U}(\bm{\theta}) = \prod_{l=1}^L \left( \bigotimes_{q=1}^n R_y(\theta_{l,q}) \cdot \mathrm{CZ}_{\text{ring}} \right).
\end{equation}
From a Lie-algebraic perspective, repetitive applications of the $R_y$ and $\mathrm{CZ}$ generator sets do not generate the full universal unitary group $\mathrm{SU}(2^n)$. Instead, they are confined within a restricted, lower-dimensional dynamical Lie algebra. Beyond a critical depth, any newly added layer of $R_y(\theta)$ rotations generates parametric gradient directions that are merely linear combinations of previously spanned directions in the state space. Consequently, dominant Fisher spectral volume saturates, and the additional parameters contribute redundant degrees of freedom, injecting negligible new volume to the underlying Fisher information manifold.

\subsection{\label{app:spectral_effdim}Spectral Origin of the Effective-Dimension Ceiling}

In our simulation, at $L=12$ ($P=52$), the plain PQC exhibits a high participation-ratio effective rank
($\mathrm{eff\_rank}\approx 49.3$). This should not be interpreted as an algebraic rank, but as a measure of how broadly the Fisher spectrum is distributed:
\begin{equation}
\mathrm{eff\_rank}=\frac{\left(\sum_{i=1}^{P}\lambda_i\right)^2}{\sum_{i=1}^{P}\lambda_i^2}.
\end{equation}
Thus, $\mathrm{eff\_rank}$ can remain large even when many directions carry only weak statistical information.
The relevant quantity for usable expressivity is instead the log-volume functional appearing in the effective dimension.

Let $F(\theta)$ denote the empirical Fisher information matrix of the circuit family at parameter point $\theta$, and let
\begin{equation}
\widehat F(\theta)=\frac{P}{\mathbb{E}_{\theta}\operatorname{Tr}F(\theta)}F(\theta)
\end{equation}
be the normalized Fisher matrix used in the effective-dimension estimator. For
\begin{equation}
\kappa(n_{\mathrm{data}})=\frac{n_{\mathrm{data}}}{2\pi\log n_{\mathrm{data}}},
\end{equation}
the finite-sample effective dimension can be written exactly as
\begin{equation}
\mathfrak{D}_N=\frac{2}{\log \kappa}\log\mathbb{E}_{\theta}\left[\sqrt{\det\left(I_P+\frac{\kappa}{2}\widehat F(\theta)\right)}\right].
\label{eq:effdim_exact_fisher}
\end{equation}
Equivalently, if $\widehat\lambda_i(\theta)$ are the eigenvalues of $\widehat F(\theta)$,
\begin{equation}
\mathfrak{D}_N=\frac{2}{\log \kappa}\log\mathbb{E}_{\theta}\exp\left[\frac12\sum_{i=1}^{P}\log\left(1+\frac{\kappa}{2}\widehat\lambda_i(\theta)\right)\right].
\label{eq:effdim_exact_spectrum}
\end{equation}
This identity shows that the ceiling is not determined by the number of parameters alone, nor by the participation-ratio effective rank alone. It is determined by the entire normalized Fisher spectrum.

The saturation observed for the plain $R_y+\mathrm{CZ}$ ansatz is therefore a spectral saturation phenomenon. Beyond a moderate depth, additional layers introduce directions whose Fisher eigenvalues are small enough that their contributions
\begin{equation}
\log\left(1+\frac{\kappa}{2}\widehat\lambda_i\right)
\end{equation}
are negligible. 
Although the parameter count $P$ and the participation-ratio effective rank may still increase, the accumulated log-volume in Eq.~\eqref{eq:effdim_exact_spectrum} stops growing.

In the depth-saturated regime, the empirical Fisher spectra vary only weakly with further increases in $L$, and the log-volume functional in Eq.~\eqref{eq:effdim_exact_spectrum} becomes nearly constant.
In the numerical evaluation, however, we do not replace the Fisher spectrum by a single analytic spectrum.
For each sampled parameter point $\theta_s$, let $\widehat\lambda_i(\theta_s)$ denote the eigenvalues of the normalized Fisher matrix $\widehat F(\theta_s)$.
The reported effective dimension is computed from the finite-sample estimator
\begin{equation}
\widehat{\mathfrak{D}}_N=\frac{2}{\log \kappa}\log\left[\frac{1}{S}\sum_{s=1}^{S}\exp\left(\frac12\sum_{i=1}^{P}\log\left(1+\frac{\kappa}{2}\widehat\lambda_i(\theta_s)\right)\right)\right].
\label{eq:effdim_finite_sample}
\end{equation}
Applying this estimator to the sampled Fisher spectra of the deepest plain PQC
at $L=12$ gives
\begin{equation}
\widehat{\mathfrak{D}}_N^{\mathrm{plain}}=25.3868\approx25.4.
\end{equation}

This value is therefore not a heuristic correction to a rank estimate, nor is it obtained by replacing the sample average with a single spectrum.
It is the finite-sample effective-dimension estimator evaluated on the normalized Fisher spectra of the plain ansatz.
The approximate invariance of these spectra in the saturated-depth regime explains why this estimator plateaus: deeper $R_y+\mathrm{CZ}$ layers fail to generate additional statistically resolvable Fisher volume, even though they continue to add formal parameters and weak spectral components.
In contrast, the proposed ansatz reshapes the Fisher spectrum by placing more weight on resolvable nonzero directions, which explains its continued growth in effective dimension under the same hardware-resource budget.

\end{document}